\documentclass[aps,prb,showpacs,twocolumn,superscriptaddress]{revtex4-2}

\usepackage{amsmath}
\usepackage{amsfonts}
\usepackage{amssymb}
\usepackage{epsfig}
\usepackage{tensor}

\usepackage{braket}

\usepackage{graphicx}

\usepackage{bbold}

\usepackage{hyphenat}
\usepackage{hyperref}
\usepackage{placeins}	
\usepackage{xcolor}
\usepackage{csquotes}

\usepackage{array}
\newcolumntype{C}[1]{>{\centering\arraybackslash}m{#1}}
\usepackage{overpic}

\usepackage[nameinlink,noabbrev]{cleveref}
\usepackage{physics}

\newcommand{\ua}{\uparrow}
\newcommand{\da}{\downarrow}

\renewcommand{\vec}{\mathbf}

\DeclareUnicodeCharacter{0301}{\'{e}}

\usepackage[normalem]{ulem}

\usepackage{booktabs}
\usepackage{diagbox}
\usepackage{siunitx}
\AtBeginDocument{
	\heavyrulewidth=.08em
	\lightrulewidth=.05em
	\cmidrulewidth=.03em
	\belowrulesep=.65ex
	\belowbottomsep=0pt
	\aboverulesep=.4ex
	\abovetopsep=0pt
	\cmidrulesep=\doublerulesep
	\cmidrulekern=.5em
	\defaultaddspace=.5em
}
\begin{document}

\title{Quantum phase transitions in the $K$-layer Ising toric code}
\author{Lukas Schamri\ss}
\affiliation{Lehrstuhl f\"ur Theoretische Physik I, Staudtstra{\ss}e 7, Universit\"at Erlangen-N\"urnberg, D-91058 Erlangen, Germany}

\author{Lea Lenke}
\affiliation{Lehrstuhl f\"ur Theoretische Physik I, Staudtstra{\ss}e 7, Universit\"at Erlangen-N\"urnberg, D-91058 Erlangen, Germany}

\author{Matthias M\"uhlhauser}
\affiliation{Lehrstuhl f\"ur Theoretische Physik I, Staudtstra{\ss}e 7, Universit\"at Erlangen-N\"urnberg, D-91058 Erlangen, Germany}

\author{Kai Phillip Schmidt}
\affiliation{Lehrstuhl f\"ur Theoretische Physik I, Staudtstra{\ss}e 7, Universit\"at Erlangen-N\"urnberg, D-91058 Erlangen, Germany}

\begin{abstract}
 We investigate the quantum phase diagram of the $K$-layer Ising toric code corresponding to $K$ layers of two-dimensional toric codes coupled by Ising interactions. While for small Ising interactions the system displays $\mathbb{Z}_2^K$ topological order originating from the toric codes in each layer, the system shows $\mathbb{Z}_2$ topological order in the high-Ising limit. The latter is demonstrated for general $K$ by deriving an effective low-energy model in $K^{\rm th}$-order degenerate perturbation theory, which is given as an effective anisotropic single-layer toric code in terms of collective pseudo-spins 1/2 refering to the two ground states of isolated Ising chain segments. For the specific cases $K=3$ and $K=4$ we apply high-order series expansions to determine the gap series in the low- and high-Ising limit. Extrapolation of the elementary energy gaps gives convincing evidence that the ground-state phase diagram consists of a single quantum critical point in the 3d Ising* universality class for both $K$ separating both types of topological order, which is consistent with former findings for the bilayer Ising toric code.
\end{abstract}

\maketitle

%Introduction
%%%%%%%%%%%%%%%%%%%%%%%%%%%%%%%%%%%%%%%%%%%%%%%%%%%%%%%%%%%%%%%%%%%%%%%%%%%%%%%%%%%%%%%%%%%%
\section{Introduction}

Two-dimensional topological order \cite{Wen_1989,Wen_1990,Wen_2004} is known for its fascinating physical properties like long-range entangled ground states, a topology-dependent ground-state degeneracy, and elementary anyonic excitations with fractional statistics \cite{Leinaas_1977,Wilczek_1982}, all related by the universal topological entanglement entropy \cite{Kitaev_2006_b,Levin_2006}. These features are further at the heart of potential applications as topological quantum computers or quantum memories \cite{Kitaev_2003,Nayak_2008}. In recent years also topological order in three-dimensional quantum many-body systems has been explored and, apart from direct generalizations of intrinsic topological order with a ground-state degeneracy depending only on the genus of the underlying topology like in the 3D toric code \cite{Hamma_2005,Nussinov_2008,Reiss_2019}, so-called fracton phases have received a lot of attention \cite{Chamon_2005,Bravyi_2011,Haah_2011,Yoshida_2013,Vijay_2015,Vijay_2016,Muehlhauser_2020,Muehlhauser_2021}. Here the ground-state degeneracy scales sub-extensively with the system size and elementary fracton excitations are immobile even in the presence of arbitrary local perturbations. Interestingly, many fracton models can originate from a layer construction \cite{Ma_2017, Vijay_2017}, i.e., two-dimensional codes exhibiting topological order are stacked in specific ways resulting in three-dimensional quantum systems with non-trivial topological properties. 

In general, it is therefore an important and fundamental question how topological order in two dimensions changes when adding a finite extension in the vertical direction so that one scales from two to three spatial dimensions. This includes also the fate of topological phase transitions out of topologically ordered phases which can not be described  by local order parameters. One promising framework for topological phase transitions in two dimensions is in terms of the condensation of bosonic quasiparticles, also dubbed topological symmetry breaking \cite{Bais_2002,Bais_2007,Bais_2009,Burnell_2011,Burnell_2018}, which has been observed microscopically in a variety of models \cite{Trebst_2007,Hamma_2008_b,Yu_2008,Vidal_2009,Vidal_2011,Dusuel_2009,Tupitsyn_2010,Wu_2012,Dusuel_2011,Schmidt_2013,Jahromi_2013,Morampudi_2014,Schulz_2016,Zhang_2017,Vanderstraeten_2017} for phase transitions between topological and non-topological phases. This concept can also be extended to phase transitions between two distinct topological phases. Apart from bilayer fractional quantum Hall systems \cite{Wen_2010,Barkeshli_2010,Moeller_2014} and certain lattice models \cite{Bombin_2008,Morampudi_2014,Schulz_2016}, the bilayer Ising toric code represents a paradigmatic example \cite{Fujii_2019,Wiedmann_2020}. The latter consists of two toric code layers coupled by an Ising interaction. The toric code (TC) \cite{Kitaev_2003} is an exactly solvable two-dimensional quantum spin model with intrinsic $\mathbb{Z}_2$ topological order and elementary excitations with mutual Abelian statistics. The bilayer Ising TC displays a second-order quantum phase transition between $\mathbb{Z}_2\times\mathbb{Z}_2$ and $\mathbb{Z}_2$ intrinsic topological order. The associated quantum phase transition can be described by the condensation of bosonic quasiparticles from both sides and it lies in the 3d Ising$^*$ universality class, which can be deduced from an exact duality mapping to the transverse-field Ising model on the square lattice \cite{Wiedmann_2020}. 

%Figure 1 - KITC
%%%%%%%%%%%%%%%%%%%%%%%%%%%%%%%%%%%%%%%%%%%%%%%%%%%%%%%%%%%%%%%%%%%%%%%%%%%%%%%%%%%%%%%%%%%%
\begin{figure}[t]
        \centering
        \includegraphics[width=\columnwidth]{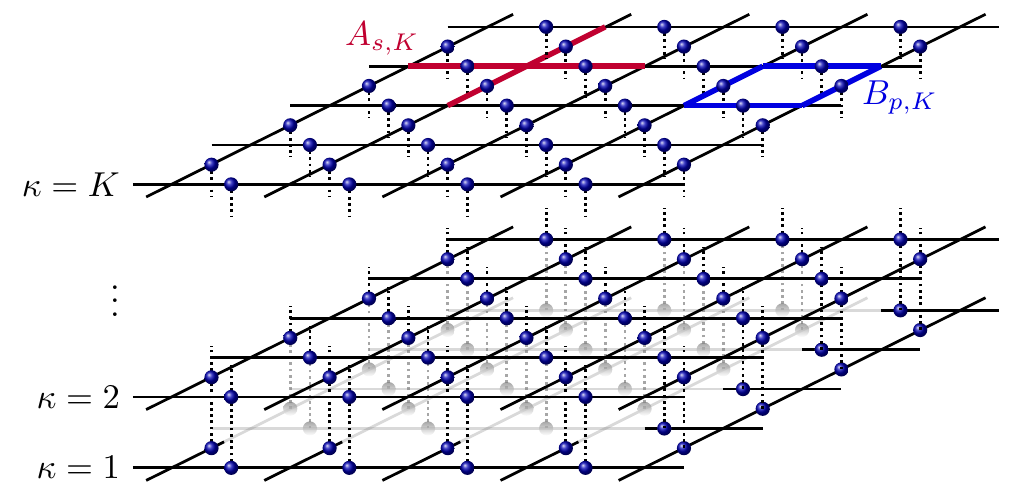}
        \caption{Illustration of the $K$-layer toric code with Ising inter-layer couplings. Blue dots represent the $N_{\rm c}$ spins in the individual $K$ layers so that the system has $N_{\rm c} K$ spins in total. The grids in each layer indicate whether a group of four spins is coupled by a star operator $A_{s,\kappa}$ or a plaquette operator $B_{p,\kappa}$ as indicated in the upper layer. The dotted lines connect spins coupled by the Ising interaction.}
        \label{Fig::KITC}
\end{figure}
%%%%%%%%%%%%%%%%%%%%%%%%%%%%%%%%%%%%%%%%%%%%%%%%%%%%%%%%%%%%%%%%%%%%%%%%%%%%%%%%%%%%%%%%%%%%

Here we study the dimensional crossover from two- to three-dimensional topological quantum systems by investigating the $K$-layer Ising toric code (KITC) corresponding to $K$ layers of two-dimensional TCs coupled by Ising interactions as illustrated in Fig.~\ref{Fig::KITC}. We aim at understanding the topological phase transition between the $\mathbb{Z}_2^K$  topological order for small Ising interactions and the $\mathbb{Z}_2$ topological order in the high-Ising limit. The presence of the topological order for large Ising interactions is demonstrated for general $K$ by deriving an effective low-energy model in $K^{\rm th}$-order degenerate perturbation theory, which is given as an effective single-layer TC. For the specific cases $K=3$ and $K=4$ we apply high-order series expansions to access the ground-state phase diagram quantitatively. We find evidence for a single quantum critical point in the 3d* universality class in both cases separating both types of topological order.

The article is organized as follows. In Sec.~\ref{sect::model} we introduce the KITC. We further describe an exact duality mapping in the low-Ising topological phase and an exact representation of the KITC in terms of pseudospins 1/2 and hardcore bosons suitable for the high-Ising limit. All results for general $K$ are contained in Sec.~\ref{sect::results_K}. This includes the derivation of an effective single-layer TC in the high-Ising limit using $K^{\rm th}$-order degenerate perturbation theory and the discussion of leading order perturbation theory for the elementary gaps in both limits. In the following Sec.~\ref{sect::results_34} we apply high-order series expansions for the specific cases $K=3$ and $K=4$ and determine the ground-state phase diagram via extrapolation of the second derivative of the ground-state energy and the gap series about both limits. Finally, we conclude the main findings and implications in Sec.~\ref{sect::conclusions}. 

%Model
%%%%%%%%%%%%%%%%%%%%%%%%%%%%%%%%%%%%%%%%%%%%%%%%%%%%%%%%%%%%%%%%%%%%%%%%%%%%%%%%%%%%%%%%%%%%
\section{Model}
\label{sect::model}
The KITC is defined on $K$ stacked two-dimensional square lattices with $N_{\rm c}$ spin-$1/2$ sites each which are located on the edges as shown in Fig.~\ref{Fig::KITC}. The spins are described by Pauli matrices $\sigma^\alpha_{j,\kappa}$ with $\alpha\in\{x,y,z\}$ where the index $\kappa\in\{ 1,\dots,K\}$ represents the layers and $j$ labels the $N_{\rm c}$ supersites which coincide with the spin sites of a single layer. These supersites correspond therefore to finite Ising chain segments with $K$ sites. The Hamiltonian
%%%%%%%%%%%%%%%%%%%%%%%%%%%%%%%%%%%%%%%%%%%%%%%%%%%%%%%%%%%%%%%%%%%%%%%%%%%%%%%%%%%%%%%%%%%%
\begin{equation} \label{HKITC}
	\mathcal{H}_\text{KITC}=-J_\text{s}\sum_{s,\kappa}A_{s,\kappa}-J_\text{p}\sum_{p,\kappa}B_{p,\kappa}-I\sum_{j,\langle \kappa, \kappa'\rangle}\sigma^z_{j,\kappa}\sigma^z_{j,\kappa'}
\end{equation}
%%%%%%%%%%%%%%%%%%%%%%%%%%%%%%%%%%%%%%%%%%%%%%%%%%%%%%%%%%%%%%%%%%%%%%%%%%%%%%%%%%%%%%%%%%%%
consists of two types of interactions. Firstly, there are four-spin-interactions within each layer defined analogously to the single layer TC as introduced by Kitaev \cite{Kitaev_2003}. Spins located around a vertex of the same layer interact according to the star operator $A_{s,\kappa}\equiv\prod_{j\in s}\sigma_{j,\kappa}^x$ where stars are centered on vertices in layer $\kappa$, whereas spins on a square in the same layer $\kappa$ interact according to the plaquette operator $B_{p,\kappa}\equiv\prod_{j\in p}\sigma_{j,\kappa}^z$.  These operators have eigenvalues $a_{s,\kappa},b_{p,\kappa}\in\{\pm 1\}$, which is a consequence of \mbox{$A^2_{s,\kappa}=B^2_{p,\kappa}=\mathbb{1}$} for all $s, p, \kappa$. Secondly, two nearest-neighbor spins in adjacent layers $\kappa$ and $\kappa'$ are coupled by an Ising interaction $\sigma^z_{j,\kappa}\sigma^z_{j,\kappa'}$. In the following we focus on $J_{\rm s},J_{\rm p},I>0$.%, since all other cases can be mapped to this one due to the fact that the TC and the Ising interaction have energy spectra which are symmetric under sign reversal. \leacomment{Reicht das als Begründung?}

Individual plaquette operators $B_{p,\kappa}$ commute with $\mathcal{H}_{\rm KITC}$ so that the $b_{p,\kappa}$ are conserved quantities. This is not the case for the eigenvalues $a_{s,\kappa}$ of the star operators $A_{s,\kappa}$ whenever $I\neq 0$, because \mbox{$[A_{s,\kappa},\sigma_{i,\kappa}^z\sigma_{i,\kappa\pm1}^z]\neq 0$} if $i\in s$. Instead, the operators $A_s^\equiv\equiv\prod_\kappa A_{s,\kappa}$ and \mbox{$B_p^\equiv\equiv\prod_\kappa B_{p,\kappa}$} commute with the Hamiltonian; we will call them superstar and superplaquette operators in the following. Whereas the $B_p^\equiv$ are not independent of the individual conserved plaquette operators, the eigenvalues $a_s^\equiv \in \{\pm 1\}$ of all $A_s^\equiv$ are additional conserved quantities. Since the possible eigenvalues of all conserved quantities found so far are $\pm1$, we refer to them as parities that divide the Hilbert space into subspaces which do not mix under the action of the Hamiltonian $\mathcal{H}_\text{KITC}$. It will turn out that the relevant low-energy physics takes place in the sector $b_{p,\kappa}=a_s^\equiv=+1$ for all $p,\kappa,s$ as for the bilayer Ising toric code \cite{Wiedmann_2020}.

%Large-Ising limit
%%%%%%%%%%%%%%%%%%%%%%%%%%%%%%%%%%%%%%%%%%%%%%%%%%%%%%%%%%%%%%%%%%%%%%%%%%%%%%%%%%%%%%%%%%%%
\subsection{Dual model: High-Ising limit}
For the discussion of the limit of high Ising interactions (hI), we apply a duality transformation that makes the symmetries of the individual Ising chain segments apparent. We observe that each chain can be described by knowing one selected spin and the alignment properties, i.e., ferro- or antiferromagnetic alingment, between all pairs of neighboring spins. Therefore, we map the $K$ spins of a chain segment at supersite $j$ on a pseudospin $\tau_j$ accounting for the $\mathbb{Z}_2$-symmetry
%%%%%%%%%%%%%%%%%%%%%%%%%%%%%%%%%%%%%%%%%%%%%%%%%%%%%%%%%%%%%%%%%%%%%%%%%%%%%%%%%%%%%%%%%%%%
\begin{align}
	\tau_j^z\equiv\sigma_{j,1}^z, && \tau_j^x\equiv \prod_{\kappa}\sigma_{j,\kappa}^x \label{eq:hI-pseudospin}
\end{align}
%%%%%%%%%%%%%%%%%%%%%%%%%%%%%%%%%%%%%%%%%%%%%%%%%%%%%%%%%%%%%%%%%%%%%%%%%%%%%%%%%%%%%%%%%%%%
and $K-1$ hardcore bosons
%%%%%%%%%%%%%%%%%%%%%%%%%%%%%%%%%%%%%%%%%%%%%%%%%%%%%%%%%%%%%%%%%%%%%%%%%%%%%%%%%%%%%%%%%%%%
\begin{align}
\begin{split}
b_{j,\beta}^{\phantom{\dagger}}&\equiv\frac{1}{2}\left(1-\sigma_{j,\beta}^z\sigma_{j,\beta+1}^z\right)\prod_{\kappa>\beta}\sigma_{j,\kappa}^x\,,\\
b_{j,\beta}^\dagger &\equiv\frac{1}{2}\left(1+\sigma_{j,\beta}^z\sigma_{j,\beta+1}^z\right)\prod_{\kappa>\beta}\sigma_{j,\kappa}^x\,,
\end{split}\label{eq:hI-bosons}
\end{align}
%%%%%%%%%%%%%%%%%%%%%%%%%%%%%%%%%%%%%%%%%%%%%%%%%%%%%%%%%%%%%%%%%%%%%%%%%%%%%%%%%%%%%%%%%%%%
where $\beta\in\{1,\ldots,K-1\}$ labels the positions between the layers numbered like the layer directly below it. An analogous mapping was used for the bilayer Ising toric code \cite{Wiedmann_2020}; it originates from a perturbative treatment of the Kitaev honeycomb model \cite{Vidal_2008}. Note that these bosonic operators $b^{\phantom{\dagger}}_{j,\beta}$ are different from the star operator eigenvalues $b_{p, \kappa}$. They can be distinguished in particular by the first index. The particle number operators are defined as $n_{j,\beta}\equiv b_{j,\beta}^\dagger b_{j,\beta}^{\phantom{\dagger}}$ and can only take the values one and zero depending on the alignment of the two spins in adjacent layers. The hardcore bosons are mutually independent, which is described by the commutation relations
%%%%%%%%%%%%%%%%%%%%%%%%%%%%%%%%%%%%%%%%%%%%%%%%%%%%%%%%%%%%%%%%%%%%%%%%%%%%%%%%%%%%%%%%%%%%
\begin{equation}
    \begin{split}
    [b_{i,\beta}^{\phantom{\dagger}},b_{j,\beta'}^{\phantom{\dagger}}]&=0,\\ [b_{i,\beta}^\dagger,b_{j,\beta'}^\dagger]&=0, \\ [b_{i,\beta}^{\phantom{\dagger}},b_{j,\beta'}^\dagger]&=\delta_{ij}\delta_{\beta\beta'}(1-2n_{i,\beta})\,.
\end{split}
\end{equation}
%%%%%%%%%%%%%%%%%%%%%%%%%%%%%%%%%%%%%%%%%%%%%%%%%%%%%%%%%%%%%%%%%%%%%%%%%%%%%%%%%%%%%%%%%%%%

In order to rewrite the original Hamiltonian (\ref{HKITC}), we need to express all $\sigma_{j,\kappa}$-matrices in terms of the pseudospin $\tau_j$ and the hardcore bosons
%%%%%%%%%%%%%%%%%%%%%%%%%%%%%%%%%%%%%%%%%%%%%%%%%%%%%%%%%%%%%%%%%%%%%%%%%%%%%%%%%%%%%%%%%%%%
\begin{align}
    &\sigma_{j,1}^x=\tau_j^x(b_{j,1}^\dagger+b_{j,1}^{\phantom{\dagger}})\,,\nonumber\\ 
    &\sigma_{j,\kappa}^x=(b_{j,\kappa-1}^\dagger+b_{j,\kappa-1}^{\phantom{\dagger}})(b_{j,\kappa}^\dagger+b_{j,\kappa}^{\phantom{\dagger}})\text{ for }1<\kappa <K\,,\nonumber\\
    &\sigma_{j,K}^x=b_{j,K-1}^\dagger+b_{j,K-1}^{\phantom{\dagger}}\,,\nonumber\\
%    &\sigma_{j,1}^y=\tau_j^y(g_{j,1}^\dagger+g_{j,1}^{\phantom{\dagger}}),\\
%    &\sigma_{j,\kappa}^y=i\tau_j^z\prod_{h=1}^{\kappa-2}(1-2n_{j,h})(g_{j,\kappa-1}^\dagger-g_{j,\kappa-1}^{\phantom{\dagger}})(g_{j,\kappa}^\dagger+g_{j,\kappa}^{\phantom{\dagger}})\text{ for }1<\kappa <K,\\
%    &\sigma_{j,K}^y=i\tau_j^z\prod_{h=1}^{K-1}(1-2n_{j,h})(g_{j,K-1}^\dagger-g_{j,K-1}^{\phantom{\dagger}}),\\
%	&\sigma^z_{j,1}=\tau_j^z,\\
	&\sigma_{j,\kappa}^z=\tau_j^z\prod_{\beta<\kappa}(1-2n_{j,\beta})\,.\label{eq:hI-paulis_from_bosons}
\end{align}
%%%%%%%%%%%%%%%%%%%%%%%%%%%%%%%%%%%%%%%%%%%%%%%%%%%%%%%%%%%%%%%%%%%%%%%%%%%%%%%%%%%%%%%%%%%%

%Figure 2 - Particles
%%%%%%%%%%%%%%%%%%%%%%%%%%%%%%%%%%%%%%%%%%%%%%%%%%%%%%%%%%%%%%%%%%%%%%%%%%%%%%%%%%%%%%%%%%%%
\begin{figure}[t]
	\centering
	\includegraphics[width=\columnwidth]{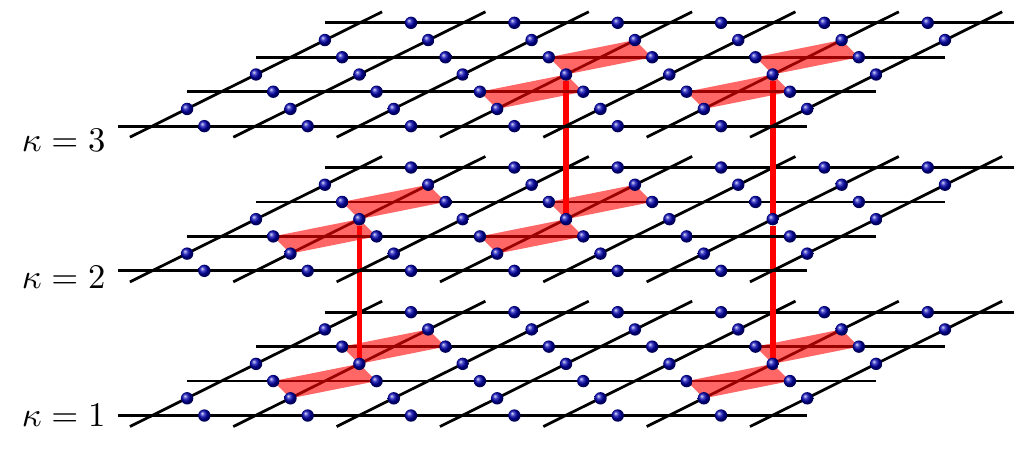}
	\caption{Illustration of charge excitations in the 3ITC in the low-Ising limit which are created locally by Ising interactions on the ground state. The spins of the original model are depicted as blue dots and are located on the edges of the lattice. Charges correspond to star operators with eigenvalue $-1$ and are illustrated as red filled squares centered at the vertices. Due to the choice of the parity sector $a_s^\equiv=+1$, any possible excitation contains either none or two charges at each superstar. The red lines connect spins whose Ising coupling creates the corresponding two pairs of charges.
		%	\leacomment{Ich habe das Gefühl die Quasiteilchen befinden sich zwischen den Layern auf Höhe der Knoten, richtig?}\lukascomment{Ja, dem würde ich zustimmen.}\kaicomment{Ich bin mir nicht sicher ob man das sagen kann. Aber es ist in jedem Fall nicht falsch!}\lukascomment{Was das Gitter betrifft, auf dem sie leben, könnte man sich ein Quadratbilayer vorstellen und das ehemalige U-3 als ein Doppelteilchen sehen, das auf zwei Plätzen gleichzeitig lebt. Ich könnte mir aber vorstellen, dass es klarer ist, von drei Teilchen in einem Singlelayer-Gitter zu sprechen, da sie alle die selbe Energie haben.}
	}
	\label{Uipart}
\end{figure}
%%%%%%%%%%%%%%%%%%%%%%%%%%%%%%%%%%%%%%%%%%%%%%%%%%%%%%%%%%%%%%%%%%%%%%%%%%%%%%%%%%%%%%%%%%%%

The so obtained Hamiltonian reads
%%%%%%%%%%%%%%%%%%%%%%%%%%%%%%%%%%%%%%%%%%%%%%%%%%%%%%%%%%%%%%%%%%%%%%%%%%%%%%%%%%%%%%%%%%%%
\begin{widetext}
\begin{equation}\label{eq:Hhb}
    \begin{split}
        \mathcal{H}_\text{dual,hI} =& -I\left[ N_{\rm c}(K-1)-2\sum_{j,\beta}n_{j,\beta}\right] -J_\text{p}\sum_p\widetilde{B}_p\sum_{\kappa}\prod_{\beta<\kappa}\prod_{j\in p}(1-2n_{j,\beta})\\
        & -J_\text{s}\sum_s\left[\widetilde{A}_s\prod_{j\in s}(b_{j,1}^\dagger+b_{j,1}^{\phantom{\dagger}})+\sum_{\beta=2}^{K-1}\prod_{j\in s}(b_{j,\beta-1}^\dagger+b_{j,\beta-1}^{\phantom{\dagger}})(b_{j,\beta}^\dagger+b_{j,\beta}^{\phantom{\dagger}})+\prod_{j\in s}(b_{j,K-1}^\dagger+b_{j,K-1}^{\phantom{\dagger}})\right]\,.
    \end{split}
\end{equation}
\end{widetext}
%%%%%%%%%%%%%%%%%%%%%%%%%%%%%%%%%%%%%%%%%%%%%%%%%%%%%%%%%%%%%%%%%%%%%%%%%%%%%%%%%%%%%%%%%%%%
where we define
%%%%%%%%%%%%%%%%%%%%%%%%%%%%%%%%%%%%%%%%%%%%%%%%%%%%%%%%%%%%%%%%%%%%%%%%%%%%%%%%%%%%%%%%%%%%
\begin{align}
    \widetilde{A}_s\equiv\prod_{j\in s}\tau^x_j\,, && \widetilde{B}_p\equiv\prod_{j\in p}\tau^z_j\,, \label{eq:hI-effective-TC-operators}
\end{align}
%%%%%%%%%%%%%%%%%%%%%%%%%%%%%%%%%%%%%%%%%%%%%%%%%%%%%%%%%%%%%%%%%%%%%%%%%%%%%%%%%%%%%%%%%%%%
in analogy to the star and plaquette operators of the TC. Interestingly, this implies the equality $A_s^\equiv=\widetilde{A}_s$ which does not hold for $B_p^\equiv$. 
%($\widetilde{B}_p=B_{p,1}$, $B_{p,\kappa}$ kompliziert)

%Low-Ising limit
%%%%%%%%%%%%%%%%%%%%%%%%%%%%%%%%%%%%%%%%%%%%%%%%%%%%%%%%%%%%%%%%%%%%%%%%%%%%%%%%%%%%%%%%%%%%
\subsection{Dual model: Low-Ising limit}
For the limiting case of independent TC layers, the perturbative starting point for the low-Ising limit (lI), we restrict the dual model to the parity sector $a_s^\equiv=b_{p,\kappa}=1$ for all $s,p,\kappa$. This is exactly the parity sector containing the ground state of the KITC in the limiting case $I=0$. We start by constructing the set of observables used to determine the ground state of the independent TC layers. The possible excitations are charges ($a_{s,\kappa}=-1$) and fluxes ($b_{p,\kappa}=-1$). We do not need to consider the fluxes since they are excluded by the choice of the parity sector $b_{p,\kappa}=+1$ for all $p,\kappa$. Charges only occur in combinations which satisfy the requirement $a_s^\equiv=+1$ on all stars which reduces the number of possible configurations. For every superstar $s$ we can find a mapping in terms of $K-1$ pseudospins $\mu_{s,\beta}^z$ located at positions $\beta\in\{1,\ldots,K-1\}$ between the layers. The eigenvalue of a pseudospin
\begin{equation}
\mu_{s,\beta}^z \doteq \prod_{\kappa>\beta}A_{s,\kappa}
\end{equation}
is $+1$ if the number of star excitations at superstar $s$ in the layers above is even and $-1$ if it is odd.

We can use these pseudospins to rewrite all operators of the original Hamiltonian \eqref{HKITC}. The star operators read
%%%%%%%%%%%%%%%%%%%%%%%%%%%%%%%%%%%%%%%%%%%%%%%%%%%%%%%%%%%%%%%%%%%%%%%%%%%%%%%%%%%%%%%%%%%%
\begin{equation}\label{def...}
\begin{split}
A_{s,1} &\doteq \mu_{s,1}^z\,,\\
A_{s,\kappa} &\doteq \mu_{s,\kappa-1}^z\mu_{s,\kappa}^z\text{ for }1<\kappa < K\,,\\
A_{s,K} &\doteq \mu_{s,K-1}^z\, .
\end{split}
\end{equation}
%%%%%%%%%%%%%%%%%%%%%%%%%%%%%%%%%%%%%%%%%%%%%%%%%%%%%%%%%%%%%%%%%%%%%%%%%%%%%%%%%%%%%%%%%%%%
A $\sigma^z_{j,\kappa}$-matrix acting on a TC eigenstate flips the eigenvalues of the two adjacent star operators in layer $\kappa$. Therefore, the action of the Ising term $\sigma^z_{j,\kappa}\sigma^z_{j,\kappa+1}$ is the combination of this process in the layers $\kappa$ and $\kappa+1$. In the pseudospin Hamiltonian this is implemented with $\mu^x_s$-matrices acting on nearest neighbors
%%%%%%%%%%%%%%%%%%%%%%%%%%%%%%%%%%%%%%%%%%%%%%%%%%%%%%%%%%%%%%%%%%%%%%%%%%%%%%%%%%%%%%%%%%%%
\begin{equation}\label{map2}
\sum_{j, \langle \kappa,\kappa'\rangle}\sigma^z_{j,\kappa}\sigma^z_{j,\kappa'} \doteq \sum_{\langle s,s'\rangle,\beta}\mu_{s,\beta}^x\mu_{s',\beta}^x\,.
\end{equation}
%%%%%%%%%%%%%%%%%%%%%%%%%%%%%%%%%%%%%%%%%%%%%%%%%%%%%%%%%%%%%%%%%%%%%%%%%%%%%%%%%%%%%%%%%%%%
In Fig.~\ref{Uipart} it is illustrated for the 3ITC how the three types of quasiparticles can be created pairwise by the Ising interaction acting on the ground state. Plugging the relations \eqref{def...} and \eqref{map2} into \eqref{HKITC} and neglecting the constant contribution of the plaquette operators, the Hamiltonian in terms of the pseudospins $\mu_{s,\kappa}$ reads
%%%%%%%%%%%%%%%%%%%%%%%%%%%%%%%%%%%%%%%%%%%%%%%%%%%%%%%%%%%%%%%%%%%%%%%%%%%%%%%%%%%%%%%%%%%%
\begin{align}
\mathcal{H}_\text{dual,lI}=& -J_\text{s}\sum_s\left(\mu^z_{s,1}+\sum_{\langle \beta,\beta'\rangle}\mu_{s,\beta}^z\mu_{s,\beta'}^z+\mu_{s,K-1}^z\right)\nonumber\\
&-I\sum_{\langle s,s'\rangle,\beta}\mu_{s,\beta}^x\mu_{s',\beta}^x\,.\label{Hpseudo}
\end{align}
%%%%%%%%%%%%%%%%%%%%%%%%%%%%%%%%%%%%%%%%%%%%%%%%%%%%%%%%%%%%%%%%%%%%%%%%%%%%%%%%%%%%%%%%%%%%
and therefore consists of Ising interactions between superimposed pseudospins and field terms on the lowermost and topmost layer.

%General K
%%%%%%%%%%%%%%%%%%%%%%%%%%%%%%%%%%%%%%%%%%%%%%%%%%%%%%%%%%%%%%%%%%%%%%%%%%%%%%%%%%%%%%%%%%%%
\section{Results for general $K$}
\label{sect::results_K}

This section includes all results which we obtained for general values of $K$. We start by deriving an effective single-layer TC using $K^{\rm th}$-order degenerate perturbation theory within the ground-state manifold of the high-Ising limit. In the following two subsections we discuss the properties of elementary excitations perturbatively up to second order about both limits.

%General K: Effective TC
%%%%%%%%%%%%%%%%%%%%%%%%%%%%%%%%%%%%%%%%%%%%%%%%%%%%%%%%%%%%%%%%%%%%%%%%%%%%%%%%%%%%%%%%%%%%
\subsection{Effective TC in high-Ising limit}

In this subsection we derive an effective low-energy model of the KITC in the high-Ising limit $J_\text{s},J_\text{p}\ll I$. The unperturbed starting point is \mbox{$J_\text{s}=J_\text{p}=0$} so that the system consists of isolated Ising chain segments. Each such supersite $j$ has two degenerate ground states corresponding to the two ferromagnetic states $\Ket{\Uparrow}_j\equiv\Ket{\uparrow\ldots\uparrow}_j$ and $\Ket{\Downarrow}_j\equiv\Ket{\downarrow\ldots\downarrow}_j$. The full system is highly degenerate and we can use degenerate perturbation theory to derive an effective low-energy model in this ground-state manifold. This is most conveniently done in the formulation \eqref{eq:Hhb}, since the unperturbed ground states are characterized by the absence of hardcore bosons (see Eq.~\eqref{eq:hI-bosons}) $n_{j,\kappa}=0$ for all $j,\kappa$. We can therefore write the effective low-energy model solely in terms of the pseudo-spins 1/2 (see Eq.~\eqref{eq:hI-pseudospin}) identifying the two eigenstates of $\tau^z_j$ with the two ferromagnetic states $\Ket{\Uparrow}_j$ and $\Ket{\Downarrow}_j$.

All unperturbed ground states have the energy \mbox{$E_{0,{\rm hI}}^{(0)}\equiv -IN_{\rm c}(K-1)$} where $N_{\rm c}$ is the number of Ising chain segments. Since plaquette operators are conserved quantities, their action in the ground-state manifold is exact in order one perturbation theory in $J_\text{p}$ while the contribution in $J_\text{s}$ vanishes. The effective low-energy model in order one then reads
%%%%%%%%%%%%%%%%%%%%%%%%%%%%%%%%%%%%%%%%%%%%%%%%%%%%%%%%%%%%%%%%%%%%%%%%%%%%%%%%%%%%%%%%%%%%
\begin{equation}
    \mathcal{H}_\text{eff,hI}^{(1)}=-KJ_\text{p}\sum_p\widetilde{B}_p-IN_{\rm c}(K-1)\, .
\end{equation}
%%%%%%%%%%%%%%%%%%%%%%%%%%%%%%%%%%%%%%%%%%%%%%%%%%%%%%%%%%%%%%%%%%%%%%%%%%%%%%%%%%%%%%%%%%%%
This is different for the perturbation $V$ proportional to $J_\text{s}$ which changes the number of hardcore bosons and therefore introduces individual spin flips in neighboring layers. In order to get a non-trivial contribution to $\mathcal{H}_\text{eff,hI}^{(K)}$ the pseudospins must be flipped and therefore all $K$ spins of a chain segment in the original formulation. As a consequence, only a trivial constant contribution can arise for perturbative orders smaller than $K$ for the KITC. The same is true for almost all perturbative contributions in order $K$ perturbation theory. In the following we are not interested in this constant energy offset $E_\text{offset}$, but we aim at calculating the non-trivial contribution in order $K$ perturbation theory analytically which will turn out to correspond to the flipping of all spins of a superstar.   

The only term in the $K^{\rm th}$ order Takahashi expansion \cite{Takahashi_1977,Klagges_2012} that does not contain intermediate $P_0$-operators reads $P_0(VS)^{K-1}VP_0$ (see also App.~\ref{sect::appendix}). Here \mbox{$P_0=\sum_j(\ketbra{\Uparrow}_j+\ketbra{\Downarrow}_j)$} denotes the projector on the ground-state manifold and \mbox{$S\equiv (1-P_0)/(E_{0,{\rm hI}}^{(0)}-\mathcal{H}_0)$} with $\mathcal{H}_0$ the unperturbed Ising part of Hamiltonian Eq.~\eqref{eq:Hhb} for \mbox{$J_\text{s}=J_\text{p}=0$}. Thus, the only contribution in the perturbative expansion up to order $K$ yielding a non-trivial contribution, which will turn out to be proportional to $\widetilde{A}_s$, can be obtained by determining all possible processes in $V^K$ that flip all spins at a superstar exactly once and calculating the eigenvalues taken by the $S$-operators for each process. The eigenvalue of $S$ is given by $(-2I\cdot4\cdot\#QP)^{-1}$ where $\#QP$ denotes the number of hardcore bosons existing in one of the four involved Ising chain segments for later convenience. The number of hardcore bosons is determined by the sequence the spins are flipped. The $K!$ possible sequences can be labeled by permutations $\sigma\in S_K$ with $S_K$ denoting the symmetric group for permutations of $K$ elements such that the $\sigma(m)$-th spin in each chain is flipped by the $m^{\rm th}$ occurring $V$. For each permutation $\sigma$ the number of hardcore bosons per chain after the action of $n$ perturbations is denoted by $\#QP(\sigma,n)$. Finally, this yields the explicit expression% for $J_\text{s}^\text{eff}$

%%%%%%%%%%%%%%%%%%%%%%%%%%%%%%%%%%%%%%%%%%%%%%%%%%%%%%%%%%%%%%%%%%%%%%%%%%%%%%%%%%%%%%%%%%%%
%\begin{equation}
%	\begin{split}
%		&P_0(VS)^{K-1}VP_0\\
%		%&=(-J_\text{s})^K\sum_s\widetilde{A}_s\sum_{\sigma\in S_K}\prod_{n=1}^{K-1}\frac{1}{-8I\#QP(\sigma,n)}P_0+E_\text{offset}P_0\\
%		&=-\underbrace{\frac{J_\text{s}^K}{(8I)^{K-1}}\sum_{\sigma\in S_K}\prod_{n=1}^{K-1}\frac{1}{\# QP(\sigma,n)}}_{=:J_\text{s}^\text{eff}}\sum_s \widetilde{A}_s P_0+E_\text{offset}P_0\,,
%	\end{split}
%\end{equation}
\begin{align}
&P_0(VS)^{K-1}VP_0\\
&=-\frac{J_\text{s}^K}{(8I)^{K-1}}\underbrace{\sum_{\sigma\in S_K}\prod_{n=1}^{K-1}\frac{1}{\# QP(\sigma,n)}}_{2^{K-1}}\sum_s \widetilde{A}_s P_0+E_\text{offset}P_0\,.\nonumber
\end{align}
%%%%%%%%%%%%%%%%%%%%%%%%%%%%%%%%%%%%%%%%%%%%%%%%%%%%%%%%%%%%%%%%%%%%%%%%%%%%%%%%%%%%%%%%%%%%
%where $J_\text{s}^\text{eff}=J_\text{s}^K/(4I)^{K-1}$ holds, because $\sum_{\sigma\in S_K}\prod_{n=1}^{K-1}\frac{1}{\# QP(\sigma,n)}=2^{K-1}$
(see App.~\ref{sect::appendix} for a proof by induction on $K$). Accordingly, the effective Hamiltonian of the KITC in the limit $I\gg J_\text{s}$ reads
%%%%%%%%%%%%%%%%%%%%%%%%%%%%%%%%%%%%%%%%%%%%%%%%%%%%%%%%%%%%%%%%%%%%%%%%%%%%%%%%%%%%%%%%%%%%
\begin{equation}
    \mathcal{H}_\text{eff,hI}^{(K)}=-\frac{J_\text{s}^K}{(4I)^{K-1}}\sum_s\widetilde{A}_s-KJ_\text{p}\sum_p\widetilde{B}_p+E_\text{offset}\,.\label{eq:hI-GS_Hamiltonian}  
\end{equation}
%%%%%%%%%%%%%%%%%%%%%%%%%%%%%%%%%%%%%%%%%%%%%%%%%%%%%%%%%%%%%%%%%%%%%%%%%%%%%%%%%%%%%%%%%%%%
This effective model corresponds to a single-layer TC with different couplings in front of star and plaquette operators, which is again exactly solvable. The ground states in the high-Ising phase are characterized by eigenvalues $\Tilde{a}_s=\Tilde{b}_p=+1$ for all $s,p$. Consequently, the KITC for large $I$ displays $\mathbb{Z}_2$ topological order with the topological entanglement entropy $\gamma_{\rm TC}=\log 2$ of a single TC \cite{Castelnovo_2008} and the topologically ordered ground states lie in the parity sector with \mbox{$a_s^\equiv =\Tilde{a}_s=+1$} for all $s$.

We note that higher-order corrections in $J/I$ are always products of effective star and plaquette operators. As a consequence, the effective low-energy model remains exactly solvable at any order in perturbation theory. This is a consequence of the reduced Hilbert space dimension $2^{N_{\rm c}/2}$ of the low-energy subspace, which is similar to the multi-plaquette expansion in the Kitaev's honeycomb model \cite{Schmidt_2008} about the anisotropic limit. 
    
%General K: Excitations
%%%%%%%%%%%%%%%%%%%%%%%%%%%%%%%%%%%%%%%%%%%%%%%%%%%%%%%%%%%%%%%%%%%%%%%%%%%%%%%%%%%%%%%%%%%%
\subsection{Excitations in the high-Ising limit}

%Figure 2 - Excitations of the high-Ising phase
%%%%%%%%%%%%%%%%%%%%%%%%%%%%%%%%%%%%%%%%%%%%%%%%%%%%%%%%%%%%%%%%%%%%%%%%%%%%%%%%%%%%%%%%%%%%
\begin{figure}[t]
        \centering
        \includegraphics[width=0.6\columnwidth]{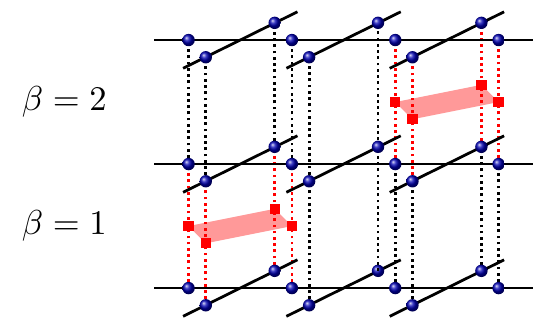}
        \caption{Illustration of the low-energy excitations in the high-Ising phase in the 3ITC. Red dotted lines mark the excited Ising bonds and red small squares indicate the corresponding hardcore bosons in the interlayers. The elementary excitations in the ground-state parity sector are the shown four hardcore-boson states.}
        \label{Fig::Excitations_Large_Ising}
\end{figure}
%%%%%%%%%%%%%%%%%%%%%%%%%%%%%%%%%%%%%%%%%%%%%%%%%%%%%%%%%%%%%%%%%%%%%%%%%%%%%%%%%%%%%%%%%%%%

In order to investigate the critical behavior, we consider the gap closing of the relevant excitations within the parity sector of the ground state that is defined by $a_s^\equiv =b_{p,\kappa}=+1$ for all $s, p, \kappa$. Here we extract the ground-state energy and the gap in second-order perturbation theory for general $K$. In Sec.~\ref{sect::results_34} we determine this gap for the specific cases $K=3$ and $K=4$ using high-order series expansions. This allows a quantitative analysis of the gap closing. Which excitations one has to consider is not a priori clear. Nevertheless, it cannot be the static excitations of the effective single-layer TC \eqref{eq:hI-GS_Hamiltonian} but hardcore bosons must be involved. From Eqs.~\eqref{eq:hI-paulis_from_bosons} and \eqref{eq:hI-effective-TC-operators} we can express
%%%%%%%%%%%%%%%%%%%%%%%%%%%%%%%%%%%%%%%%%%%%%%%%%%%%%%%%%%%%%%%%%%%%%%%%%%%%%%%%%%%%%%%%%%%%
\begin{equation}
	\begin{split}
		A_s^\equiv&=\widetilde{A}_s\,,\\
		B_{p,1}&=\widetilde{B}_p\,,\\
		B_{p,\kappa}&=\widetilde{B}_p\prod_{\beta<\kappa} \prod_{j\in p}(1-2n_{j,\beta})\,.
	\end{split}
\end{equation}
%%%%%%%%%%%%%%%%%%%%%%%%%%%%%%%%%%%%%%%%%%%%%%%%%%%%%%%%%%%%%%%%%%%%%%%%%%%%%%%%%%%%%%%%%%%%
This implies that the effective star and plaquette operators have eigenvalue $+1$ in this parity sector. Investigating terms as $B_{p,\beta}B_{p,\beta+1}=\prod_{j\in p}(1-2n_{j,\beta})$ yields that the hardcore bosonic excitations in the interlayer $\beta$ have to be distributed along closed loops on the dual lattice. 
Thus, the relevant low-energy excitations have the energy $8I$ and are of the form $\ket{s,\beta}\equiv\prod_{j\in s}g_{j,\beta}^\dagger\ket{0}$, where $\ket{0}$ denotes the ground state of the effective single-layer TC \eqref{eq:hI-GS_Hamiltonian}.

First, we determine the absolute value of the ground-state energy $E_{0,{\rm hI}}^{(2)}$ for general $K$ in second-order perturbation theory. It is given by
%%%%%%%%%%%%%%%%%%%%%%%%%%%%%%%%%%%%%%%%%%%%%%%%%%%%%%%%%%%%%%%%%%%%%%%%%%%%%%%%%%%%%%%%%%%%
\begin{align}
	E^{(2)}_\text{0,hI} 
	%&= E_0^{(0)} P_0 + P_0 V P_0 + P_0 V S V P_0\\
	&= E_{0,{\rm hI}}^{(0)} - \frac{N_{\rm c}K}{2} J_\text{p} - \frac{N_{\rm c} (K+2)}{32} \frac{J_\text{s}^2}{I}\, .
\end{align}
%%%%%%%%%%%%%%%%%%%%%%%%%%%%%%%%%%%%%%%%%%%%%%%%%%%%%%%%%%%%%%%%%%%%%%%%%%%%%%%%%%%%%%%%%%%%

To calculate the gap of this elementary excitation consisting of four hardcore bosons, we introduce the Fourier transformed states $\ket{(q,p),\beta}$ where $(q,p)$ denotes the momentum coordinate within the layers, normalized by a factor of $2\pi/\sqrt{N_{\rm c}/2}$ with $\sqrt{N_{\rm c}/2}$ being the width of the lattice. The effective single-particle Hamiltonian reads
%%%%%%%%%%%%%%%%%%%%%%%%%%%%%%%%%%%%%%%%%%%%%%%%%%%%%%%%%%%%%%%%%%%%%%%%%%%%%%%%%%%%%%%%%%%%
\begin{equation}
	\mathcal{H}^{(2)}_\text{1p,hI} = E_1^{(0)} P_1 + P_1 V P_1 + P_1 V S_\text{1p} V P_1\, ,
\end{equation}
%%%%%%%%%%%%%%%%%%%%%%%%%%%%%%%%%%%%%%%%%%%%%%%%%%%%%%%%%%%%%%%%%%%%%%%%%%%%%%%%%%%%%%%%%%%%
where $P_1$ refers to the projector on the single-particle (1p) Hilbert space sector and $S_\text{1p}$ refers to the corresponding resolvent with \mbox{$E_1^{(0)}\equiv E_{0,\text{hI}}^{(0)}+8I$}. The associated matrix of dimension $K-1$ from this effective Hamiltonian reads
%%%%%%%%%%%%%%%%%%%%%%%%%%%%%%%%%%%%%%%%%%%%%%%%%%%%%%%%%%%%%%%%%%%%%%%%%%%%%%%%%%%%%%%%%%%%
\begin{equation}
	\begin{split}
		&\bra{(q,p),\beta}(\mathcal{H}^{(2)}_\text{1p,hI}-E_{0,{\rm hI}}^{(2)})\ket{(q,p),\beta'}=\\
		&=8I\delta_{\beta\beta'}-\begin{pmatrix}
			f(q,p)\frac{J_\text{s}^2}{I}   & J_\text{s}                   & \frac{J_\text{s}^2}{16I}     & 0     &          & \\
			J_\text{s}                     & \frac{J_\text{s}^2}{24I}    & J_\text{s}                   & \frac{J_\text{s}^2}{16I}&   \ddots  & \\
			\frac{J_\text{s}^2}{16I}       & J_\text{s}                   & \frac{J_\text{s}^2}{24I}    & \ddots&  \ddots   & 0\\
			0                       &\frac{J_\text{s}^2}{16I}      & \ddots                &\ddots &  \ddots   &\frac{J_\text{s}^2}{16I}  \\ 
			&\ddots                 &\ddots                 &\ddots& \frac{J_\text{s}^2}{24I} & J_\text{s}\\
			&                       & 0                     &\frac{J_\text{s}^2}{16I}  &J_\text{s}    & f(q,p)\frac{J_\text{s}^2}{I}
		\end{pmatrix}_{\beta\beta'}\\
		&=:8I\delta_{\beta\beta'}-M_{\beta\beta'}
	\end{split}
\end{equation}   
%%%%%%%%%%%%%%%%%%%%%%%%%%%%%%%%%%%%%%%%%%%%%%%%%%%%%%%%%%%%%%%%%%%%%%%%%%%%%%%%%%%%%%%%%%%%
with $f(q,p)\equiv\frac{13}{48}+\frac{1}{4}\cos q+\frac{1}{4}\cos p$ which can only be diagonalized explicitly in first order for arbitrary $K$. Using the diagonalization of tridiagonal Toeplitz matrices \cite{Gover_1994}, we find
%%%%%%%%%%%%%%%%%%%%%%%%%%%%%%%%%%%%%%%%%%%%%%%%%%%%%%%%%%%%%%%%%%%%%%%%%%%%%%%%%%%%%%%%%%%%
\begin{equation}
	\Delta^{(1)}_{{\rm hI},K\ge2}=8I-2J_\text{s}\cos\frac{\pi}{K}.
\end{equation}
%%%%%%%%%%%%%%%%%%%%%%%%%%%%%%%%%%%%%%%%%%%%%%%%%%%%%%%%%%%%%%%%%%%%%%%%%%%%%%%%%%%%%%%%%%%%
in first order. In second order, it can be shown that the eigenvalue is minimal for the case $(q,p)=0$ which is also reasonable physically since the hopping processes within the layers with the negative coefficient $-J_\text{s}^2/(8I)$ should lower the energy of a delocalized state with the same eigenvalue on all stars most.

Finally, we can analyze the limiting case $K\rightarrow\infty$ by introducing periodic boundary conditions since this neglects effects of the outermost layers. The resulting matrix $M_\text{periodic}$ reads
%%%%%%%%%%%%%%%%%%%%%%%%%%%%%%%%%%%%%%%%%%%%%%%%%%%%%%%%%%%%%%%%%%%%%%%%%%%%%%%%%%%%%%%%%%%%
\begin{equation}
	\begin{split}
	M_\text{periodic}=\sum_\kappa\bigg(&\frac{J_\text{s}^2}{24I}\ketbra{(q,p),\beta}\\
	&+J_\text{s}(\ketbra{(q,p),\beta+1}{(q,p),\beta} +{\rm h.c.})\\
	&+\frac{J_\text{s}^2}{16I}(\ketbra{(q,p),\beta+2}{(q,p),\beta}+{\rm h.c.}) \bigg)\,.
\end{split}
\end{equation}
%%%%%%%%%%%%%%%%%%%%%%%%%%%%%%%%%%%%%%%%%%%%%%%%%%%%%%%%%%%%%%%%%%%%%%%%%%%%%%%%%%%%%%%%%%%%
$M_\text{periodic}$ is diagonal when Fourier transforming additionally in the direction within the chains. The maximal eigenvalue is $2J_\text{s}+{J_\text{s}^2}/(6I)$ at $(q,p)=0$, which suffices to deduce the energy gap
\begin{equation}
	\Delta_{{\rm hI},K=\infty}^{(2)}=8I-2J_\text{s}-\frac{1}{6}\frac{J_\text{s}^2}{I}
\end{equation}
in the limit $K\rightarrow\infty$.

%General K: Excitations
%%%%%%%%%%%%%%%%%%%%%%%%%%%%%%%%%%%%%%%%%%%%%%%%%%%%%%%%%%%%%%%%%%%%%%%%%%%%%%%%%%%%%%%%%%%%
\subsection{Excitations in the low-Ising limit}

We turn to the low-Ising limit and determine the ground-state energy and the energy gap in second-order perturbation theory for general $K$. The ground state energy $E_{0,\text{lI}}^{(2)}$ is given by
%%%%%%%%%%%%%%%%%%%%%%%%%%%%%%%%%%%%%%%%%%%%%%%%%%%%%%%%%%%%%%%%%%%%%%%%%%%%%%%%%%%%%%%%%%%%
\begin{equation}
	E_{0,\text{lI}}^{(2)} = E_{0,\text{lI}}^{(0)} - \frac{N_{\rm c}(K-1)}{8} \frac{I^2}{J_\text{s}}\,.
\end{equation}
%%%%%%%%%%%%%%%%%%%%%%%%%%%%%%%%%%%%%%%%%%%%%%%%%%%%%%%%%%%%%%%%%%%%%%%%%%%%%%%%%%%%%%%%%%%%
As for the high-Ising phase, the energy gap is located at momentum $(q,p)=0$. In second-order perturbation theory, one has to distinguish the cases $K=3$ and $K\geq 4$. For the 3ITC, one has the three elementary excitations in terms of the pseudospin model Eq.~\eqref{Hpseudo} (see Fig.~\ref{Uipart} for illustration in the original language of star eigenvalues): $\ket{\ua\da}$, $\ket{\da\ua}$, and $\ket{\ua\ua}$ above the ground-state configuration $\ket{\da\da}$. The three corresponding energy gaps are given by
%%%%%%%%%%%%%%%%%%%%%%%%%%%%%%%%%%%%%%%%%%%%%%%%%%%%%%%%%%%%%%%%%%%%%%%%%%%%%%%%%%%%%%%%%%%%
\begin{equation}
\begin{split}
\Delta_{{\rm lI},\ket{\ua\da}}^{(2)}= \Delta_{{\rm lI},\ket{\da\ua}}^{(2)} &=4J_\text{s}-4I-\frac{3}{2}\frac{I^2}{J_\text{s}}\\
\Delta_{{\rm lI},\ket{\ua\ua}}^{(2)}&=4J_\text{s}-3\frac{I^2}{J_\text{s}}\, .
\end{split}
\end{equation}
%%%%%%%%%%%%%%%%%%%%%%%%%%%%%%%%%%%%%%%%%%%%%%%%%%%%%%%%%%%%%%%%%%%%%%%%%%%%%%%%%%%%%%%%%%%%
The fact that matrix elements between different particles vanish due to the conserved parity $\prod_s\mu_{s,\kappa}^z$ for each layer $\kappa$, is used for the result. For $K\ge 4$, the excitation with the lowest energy can be identified in general from this. For $I=0$ these are all states with exactly two $A_{s,\kappa}$ being flipped on a single star $s$ which results in the excitation energy $4J_\text{s}$. From the gap of the particles $\ket{\ua\da}$ and $\ket{\da\ua}$ as compared to the particle $\ket{\ua\ua}$ it can be concluded that all particles with non-vanishing first-order corrections have the flipped eigenvalues in neighboring layers. These excitations take the following form in the pseudospin model in terms of $(K-1)$-mer states on a superstar $s$: $\ket{\ua\da\dots\da}=:\ket{s,1}_\text{lI}$, $\ket{\da\ua\dots\da}=:\ket{s,2}_\text{lI}$, $\dots$, $\ket{\da\da\dots\ua}=:\ket{s,K-1}_\text{lI}$.
%%%%%%%%%%%%%%%%%%%%%%%%%%%%%%%%%%%%%%%%%%%%%%%%%%%%%%%%%%%%%%%%%%%%%%%%%%%%%%%%%%%%%%%%%%%%
%\begin{equation*}
%\ket{\ua\da\dots\da}=:\ket{1,s},\;\ket{\da\ua\dots\da}=:\ket{2,s},\;\dots,\;\ket{\da\da\dots\ua}=:\ket{K-1,s}\,.
%\end{equation*}
%%%%%%%%%%%%%%%%%%%%%%%%%%%%%%%%%%%%%%%%%%%%%%%%%%%%%%%%%%%%%%%%%%%%%%%%%%%%%%%%%%%%%%%%%%%%
For all these particles the first-order correction
%%%%%%%%%%%%%%%%%%%%%%%%%%%%%%%%%%%%%%%%%%%%%%%%%%%%%%%%%%%%%%%%%%%%%%%%%%%%%%%%%%%%%%%%%%%%
\begin{equation}
	\tensor*[_{\text{lI}}]{\bra{(q,p),\beta}}{}	P_1VP_1\ket{(q,p),\beta}_\text{lI}=-2I\left(\cos q+\cos p\right)\,,
\end{equation}
where $V$ corresponds to the Ising interaction, $P_1$ is the projector onto the single particle space and $\ket{(q,p),\beta}_\text{lI}$ denotes the Fourier transformed states.
%%%%%%%%%%%%%%%%%%%%%%%%%%%%%%%%%%%%%%%%%%%%%%%%%%%%%%%%%%%%%%%%%%%%%%%%%%%%%%%%%%%%%%%%%%%%
In second order, one finds that the energy gap is given by
%%%%%%%%%%%%%%%%%%%%%%%%%%%%%%%%%%%%%%%%%%%%%%%%%%%%%%%%%%%%%%%%%%%%%%%%%%%%%%%%%%%%%%%%%%%%
\begin{equation}
	\Delta_{{\rm lI},K\geq4}^{(2)}=4J_\text{s}-4I-2\frac{I^2}{J_\text{s}}
\end{equation}
%%%%%%%%%%%%%%%%%%%%%%%%%%%%%%%%%%%%%%%%%%%%%%%%%%%%%%%%%%%%%%%%%%%%%%%%%%%%%%%%%%%%%%%%%%%%
is determined by the quasiparticles that are not in the layers at the border.

\section{Discussion for $K = 3$ and $K = 4$}
\label{sect::results_34}

After having discussed the properties of the KITC for general $K$, we use high-order series expansions for $K = 3$ and $K = 4$ in order to gain quantitative insights in the ground-state phase diagram. We applied Löwdin's partioning technique \cite{Loewdin1962, Yao2000, Kalis_2012} and the method of perturbative continuous unitary transformations (pCUTs) \cite{Knetter_2000, Knetter2003} to calculate the ground-state energies $E_{0,\rm lI}$ and $E_{0,\rm hI}$ as well as the elementary excitation gaps $\Delta_{\rm lI}$ and $\Delta_{\rm hI}$ in the low- and high-Ising limit up to high orders in perturbation. We reached order 14 for the ground-state energy in all cases while order 11 (order 8) has been calculated for $\Delta_{\rm lI}$ ($\Delta_{\rm hI}$) for both values of $K$. Assuming a second-order phase transition, we expect the following behaviour close to the quantum critical point
%%%%%%%%%%%%%%%%%%%%%%%%%%%%%%%%%%%%%%%%%%%%%%%%%%%%%%%%
\begin{equation}
	\Delta\propto(x-x_{\rm c})^{\nu z}\text{ for }|x-x_{\rm c}|\ll1
\end{equation}
%%%%%%%%%%%%%%%%%%%%%%%%%%%%%%%%%%%%%%%%%%%%%%%%%%%%%%%%
with the critical exponents $\nu$ describing the correlation length and the dynamical critical exponent $z$ describing the autocorrelation time. The second derivative of the ground-state energy $E_0$ diverges at the critical point as follows 
%%%%%%%%%%%%%%%%%%%%%%%%%%%%%%%%%%%%%%%%%%%%%%%%%%%%%%%%
\begin{equation}
	\frac{d^2 E_0}{d x^2}\propto(x-x_{\rm c})^{-\alpha}\text{ for }|x-x_{\rm c}|\ll1,
\end{equation}
%%%%%%%%%%%%%%%%%%%%%%%%%%%%%%%%%%%%%%%%%%%%%%%%%%%%%%%%
with the critical exponent $\alpha$. Here $x$ corresponds to $J/I$ ($I/J$) in the low-Ising (high-Ising) phase.  The obtained series are extrapolated using DLog Pad\'{e} extrapolation \cite{Guttmann_1989} in order to extract estimates for the critical points $x_{\rm c}$ as well as the associated critical exponents $z\nu$ and $\alpha$. Technical details on the Löwdin's partioning technique, on the pCUT method, the explicit series as well as on DLog Pad\'{e} extrapolation can be found in App.~\ref{app_technical_aspects}.

Although we can not exclude a first-order phase transition between the low-Ising and the high-Ising phase, we can at least check consistency between the extrapolation of the low-Ising and the high-Ising series expansions, e.g., a first-order transition would result in a crossing of the two ground-state energies without a divergence of the second derivative. In principle, the ground-state phase diagram could also consist of intermediate phases. If the transitions from the low- and high-Ising phase to the intermediate region are both continuous, we could detect an intermediate region by the gap closing from both limits. In contrast, first-order phase transitions to potential intermediate phases can not be seen by our series expansions. In the following we find convincing indications for a single second-order phase transition in the 3d* universality class for $K=3$ and $K=4$. 

%Phase transition
%%%%%%%%%%%%%%%%%%%%%%%KITC%%%%%%%%%%%%%%%%%%%%%%%%%%%%%%%%%%%%%%%%%%%%%%%%%%%%%%%%%%%%%%%%%%%%%
\subsection{Phase diagram for $K = 3$}

\begin{figure*}
	\centering
	\includegraphics[width=\textwidth]{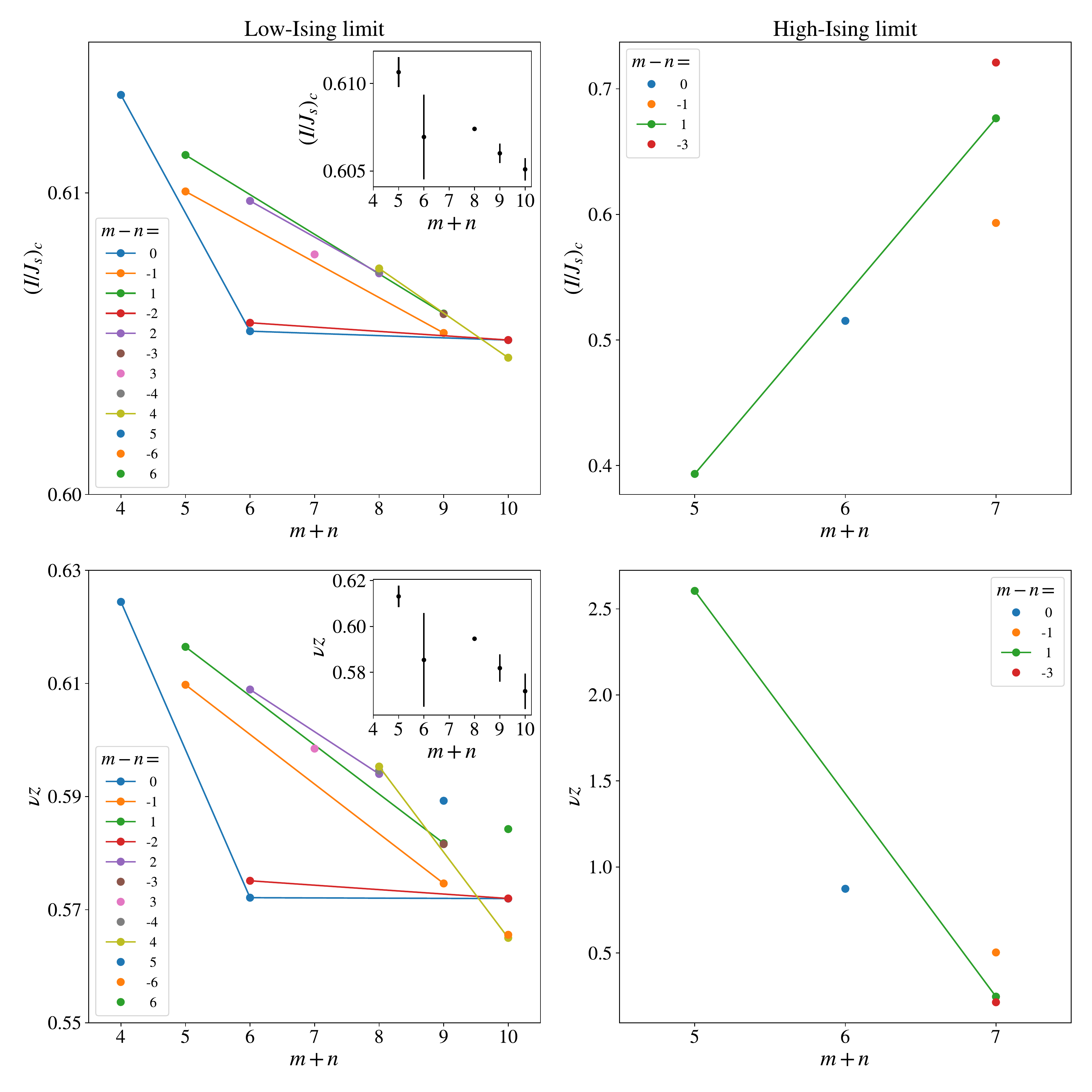}
	\caption{Non-defective DLog Padé approximants $[m,n]$ for the critical point (upper panels) and the critical exponent $\nu z$ (lower panels) for the low-Ising (left panels) and the high-Ising (right panels) limit from the $K=3$ gap series as a function of $m+n$. DLog Padé approximants $[m,n]$ of the same family with $m-n$ constant are connected by solid lines. For the more reliable gap $\Delta_\text{lI}$, additionally, the average of orders with more than one DLog Padé approximant is shown in the inset plots with the sample standard deviation as error bars.}
	\label{PadeGapK3}
\end{figure*}

\begin{figure*}
	\centering
	\includegraphics[width=\textwidth]{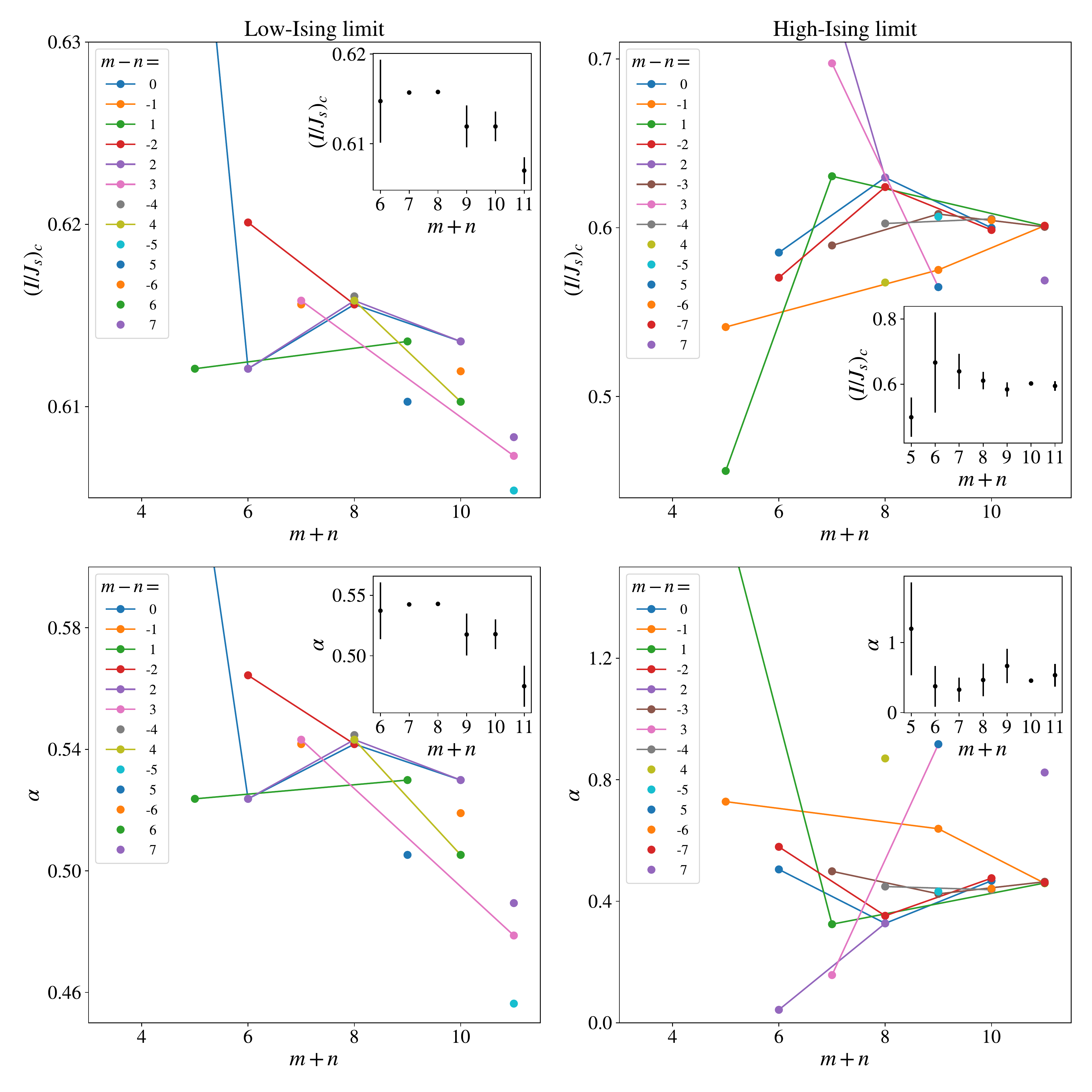}
	\caption{Non-defective DLog Padé approximants $[m,n]$ for the critical point (upper panels) and the critical exponent $\alpha$ (lower panels) for the low-Ising (left panels) and the high-Ising (right panels) limit from the second derivative of the $K=3$ ground-state energy series as a function of $m+n$. DLog Padé approximants $[m,n]$ of the same family with $m-n$ constant are connected by solid lines. The average of orders is shown in the inset plots with the sample standard deviation as error bars.}
	\label{PadeGSK3}
\end{figure*}

\renewcommand{\arraystretch}{1.2}
\setlength{\tabcolsep}{12pt} % General space between cols (6pt standard)
\begin{table}[tb]
	\centering
	\begin{tabular}{l|lll}
		&$(I/J_\text{s})_{\rm c}$ & $\nu z$ & $\alpha$ \\\hline
		$\Delta_{\rm lI}$ &$0.6056(10)$ &$0.577(11)$ &\\
		$\Delta_{\rm hI}$ &$0.63(10)$ &$0.5(4)$    &\\
		$\partial_{I/J_\text{s}}^2E_{0,\rm lI}$  &$0.6112(10)$ & &$0.504(11)$\\
		$\partial_{I/J_\text{s}}^2E_{0,\rm hI}$  &$0.6011(23)$ & &$0.461(13)$
		
	\end{tabular}
	\caption{Estimates for the critical point and the critical exponents from DLog Padé extrapolations for 3ITC. Except for the high-Ising gap, the respective values are obtained by averaging over DLog Padé approximants $[m,n]$ of the two highest orders, excluding DLog Padé families consisting of single approximants. A family consists of all approximants with equal $m-n$. The standard deviations are given in brackets. Because of the low number of approximants of the high-Ising gap $\Delta_{\rm hI}$, we also took the respective single family members into account; the approximants must be assumed to be not satisfyingly converged. Averaging over the other three results for the critical point overall indicates an estimate of $(I/J_\text{s})_{\rm c}=0.606(6)$.} 
	\label{tabelK3}
\end{table}

We start with the case \mbox{$K=3$}. The estimates for $(I/J_{\rm s})_{\rm c}$ and $z\nu$ from the gap extrapolations are displayed in Fig.~\ref{PadeGapK3}. The low-Ising gap \eqref{GapLIK3} is available up to order 11 while the high-Ising gap \eqref{GapHIK3} has been calculated up to order 8. Thus the number of DLog Padé approximants is considerably larger in the low-Ising case. The degree of convergence for the low-Ising case is remarkably good for the critical point and trustworthy for the exponent. The respective numerical values are obtained by averaging over DLog Padé approximants $[m,n]$ of the two highest orders, excluding DLog Padé families consisting of single approximants. A family consists of all approximants with equal $m-n$. The results are given in the first line of Tab.~\ref{tabelK3}.

From the high-Ising gap we can extract no reliable information about the exponent and only a rough estimate of the critical point. Therefore we will not take these results into account in the following. For completeness, the second line of Tab.~\ref{tabelK3} contains the numerical values obtained analogously for the high-Ising gap series. Due to the low number of non-defective approximants, we also took families consisting of single members into account.
	
In Fig.~\ref{PadeGSK3} estimates for $x_{\rm c}$ and $\alpha$ from the extrapolations of the second-derivative of the ground-state energy are shown. The low-Ising ground-state energy \eqref{GSLIK3} and the high-Ising ground-state energy \eqref{GSHIK3} are both available up to order 14. Due to this many approximants are available and the families seem to be reliably converged. The expected accuracy of the critical point is higher than for the exponent. Numerical values are obtained by averaging over the highest order result of all families reaching up to one of the two highest orders. Families consisting of single points are excluded since no degree of convergence can be addressed. The values are stated in the lower two lines of Tab.~\ref{tabelK3}.

Comparing Figs.~\ref{PadeGapK3} and \ref{PadeGSK3}, a descending trend in all estimates in the low-Ising limit can be observed. Therefore, these results may overestimate the higher order predictions slightly. From the ground-state approximants of the high-Ising limit no trend can be extracted. Note that the real uncertainties are larger than the standard deviation as they do not contain the systematic errors from the missing higher orders. By averaging over the three well-converged results for the critical point, an overall estimate of $(I/J_\text{s})_{\rm c}=0.606(6)$ can be obtained with the standard deviation being a lower bound for the real uncertainty. The consistence of estimated critical points from the ground-state energies and the gaps is the key indication for a single second-order phase transition.

The results for the exponents can be used to classify the phase transition. Our working hypothesis from the findings for the bilayer Ising toric code \cite{Wiedmann_2020} is that it is of the 3d Ising* universality class with $\nu z = 0.629971(4)$ \cite{Pfeuty_1971, Kos_2016} and $\alpha = 0.110087(12)$ \cite{Kos_2016}. The only reliable value obtained for the gap exponent $\nu z$ is $0.577(11)$, which is in accordance to our working hypothesis of a 3d Ising* universality class. Our predictions for the critical exponent $\alpha$ from both limits are not in accordance with the 3d Ising* critical exponent $\alpha$. This is, however, a more often occurring phenomenon and was for example also reported for the high-temperature expansion of the (2+1) dimensional Ising model \cite{He_1990}. In summary, numerical indications of a single second-order quantum phase transition have been found for the critical point $I/J_\text{s}=0.606(6)$. The critical exponent $z\nu$ of the energy gap points towards the classification as a phase transition of the 3d Ising* universality class.

%Phase transitions
%%%%%%%%%%%%%%%%%%%%%%%%%%%%%%%%%%%%%%%%%%%%%%%%%%%%%%%%%%%%%%%%%%%%%%%%%%%%%%%%%%%%%%%%%%%%
\subsection{Phase diagram for $K = 4$}
\begin{figure*}
	\centering
	\includegraphics[width=\textwidth]{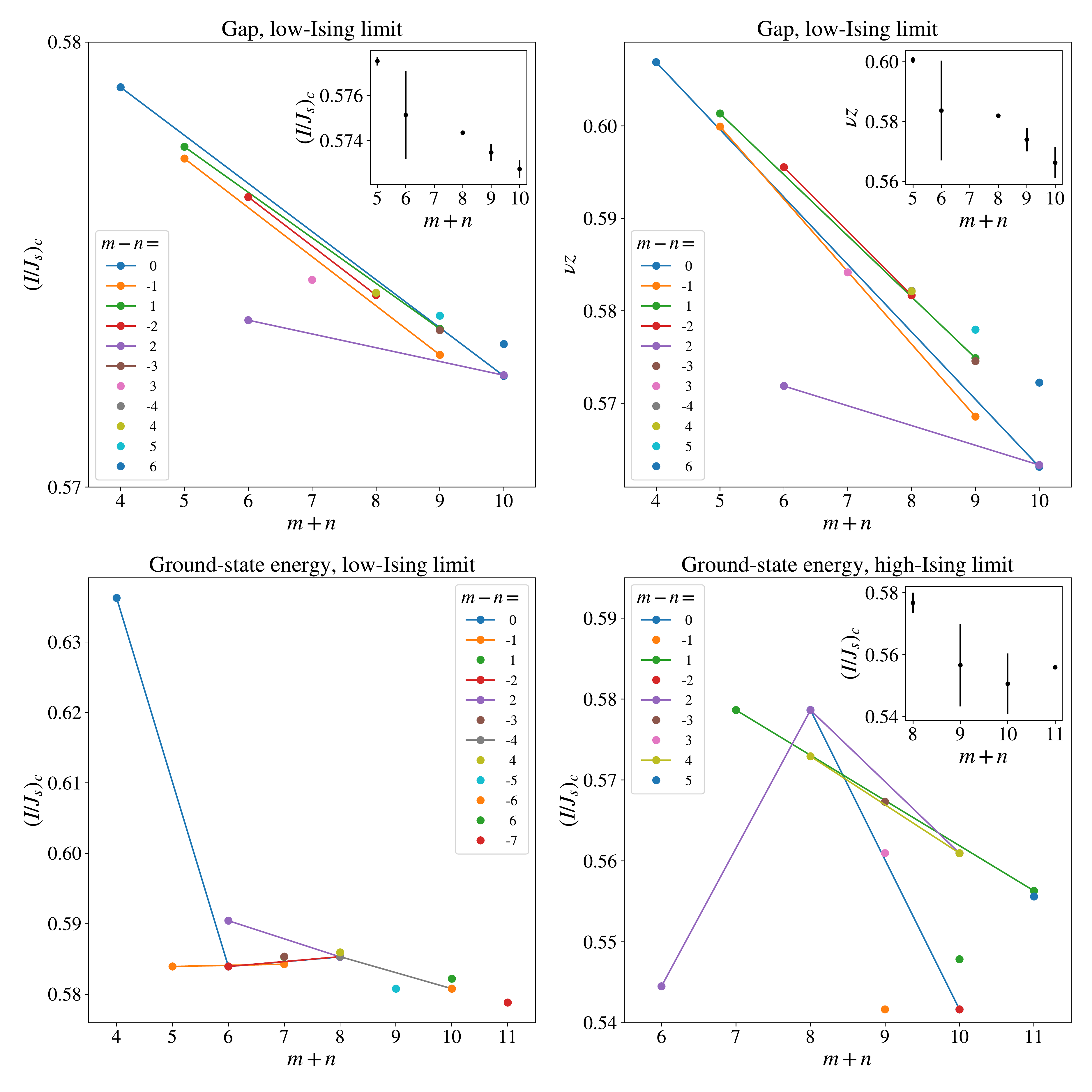}
	\caption{Non-defective DLog Padé approximants $[m,n]$ for the $K=4$ low-Ising gap series (upper panels), the $K=4$ low-Ising ground-state energy series (lower left panel) and the second derivative of the $K=4$ high-Ising ground-state energy series (lower right panel) as a function of $m+n$. DLog Padé approximants $[m,n]$ of the same family with $m-n$ constant are connected by solid lines. The average of orders with more than one DLog Padé approximant is shown in the inset plots with the sample standard deviation as error bars.}
	\label{PadeK4}
\end{figure*}

\renewcommand{\arraystretch}{1.2}
\setlength{\tabcolsep}{12pt} % General space between cols (6pt standard)
\begin{table}[tb]
	\centering
	\begin{tabular}{l|ll}
		&lI-limit 	&hI-limit \\\hline
		Energy gap 							&$0.5729(5)$&    \\
		Ground-state energy  				&$0.5806(14)$& $0.553(11)$
		
	\end{tabular}
	\caption{Estimates for the critical point with sample standard deviations of all reliable expansions for $K=4$. Except for the second derivative of the low-Ising ground-state energy, the respective values are obtained by averaging over DLog Padé approximants $[m,n]$ of the two highest orders, excluding families consisting of single approximants. A family consists of all approximants with equal $m-n$. The standard deviations are given in brackets. Because of the low number of high-order approximants of the low-Ising ground-state energy $E_{0,\rm lI}$, we also took the respective single family members into account. Averaging over the all three results for the critical point overall indicates an estimate of $(I/J_\text{s})_{\rm c}=0.569(15)$.}
	\label{tabelK4}
\end{table}

Next, we discuss the case \mbox{$K=4$} analogously to \mbox{$K=3$}. We only consider series expansions that seem promising, i.e., the low-Ising gap for $(I/J_{\rm s})_{\rm c}$ and $z\nu$ and the second derivative of the ground-state energies about both limits for $(I/J_{\rm s})_{\rm c}$. All respective values are plotted in Fig.~\ref{PadeK4}. The corresponding numerical values for the critical point are stated in Tab.~\ref{tabelK4}, obtained as before. For the low-Ising ground-state energy results, families with single family members were also taken into account.

As before, the low-Ising gap seems reliably converged and is subject to a downwards trend. In comparison, the ground-state approximants do not behave well. The approximants about the low-Ising limit seem converged, but keep in mind that a lot of approximants are defective. The values indicated by the high-Ising ground-state energies are more scattered. Nevertheless, the three results agree with each other. By averaging over them, an overall estimate of $(I/J_\text{s})_{\rm c}=0.569(15)$ can be obtained. The low-Ising gap estimates the critical exponent $\nu z$ to be $\nu z=0.567(6)$, which is further off the 3d Ising* critical exponent $\nu z = 0.629971(4)$ \cite{Pfeuty_1971, Kos_2016}. Nevertheless, it is close enough not to contradict our working hypothesis.

%In summary, we strongly expect a single second-order quantum phase transition, that might still be of the 3d Ising* universality class.

%Conclusions
%%%%%%%%%%%%%%%%%%%%%%%%%%%%%%%%%%%%%%%%%%%%%%%%%%%%%%%%%%%%%%%%%%%%%%%%%%%%%%%%%%%%%%%%%%%%
\section{Conclusion}
\label{sect::conclusions}
In this work, we have investigated the quantum phase diagram of the $K$-layer Ising toric code. The system displays $\mathbb{Z}_2^K$ topological order for small Ising interactions originating from the toric codes in each layer. In contrast, the system shows $\mathbb{Z}_2$ topological order in the high-Ising limit. This can be shown for general $K$ by deriving an effective low-energy model in $K^{\rm th}$-order degenerate perturbation theory. Up to an unimportant energy offset, the low-energy model corresponds to an effective single-layer toric code in terms of collective pseudo-spins 1/2 refering to the two ground states of isolated Ising chain segments. The prefactors of effective star and plaquette operators are highly anisotropic. While the effective plaquette operators are present in first-order perturbation theory, the effective star operators arise in order $K$ perturbation theory. As a consequence, the effective charge gap reduces with increasing $K$ and vanishes for $K\rightarrow\infty$.   

We further analyzed the nature of the quantum phase transition between the low-Ising and high-Ising topological orders. Our results are consistent with a single quantum critical point in the 3d Ising* universality class for all $K$ generalizing former findings for the bilayer Ising toric code \cite{Wiedmann_2020}. For the specific cases $K=3$ and $K=4$ we applied high-order series expansions to determine the series of the ground-state energy and the elementary gap in the low- and high-Ising limit. Extrapolation of the elementary energy gaps gives indeed convincing evidence that the ground-state phase diagram consists of a single quantum critical point for both $K$. The extracted critical exponents are in agreement with the 3d Ising* universality class, but the quality of the extrapolation is not sufficient for quantitative predictions. Further numerical studies are therefore needed which we leave for future research. 
\section*{Acknowledgments}
KPS acknowledges financial support by the German Science Foundation (DFG) through the grant SCHM 2511/11-1

\FloatBarrier

\begin{appendix}
	\section{Proof of the effective coupling of the KITC}
	\label{sect::appendix}
	In the following we show \mbox{$J_\text{s}^\text{eff}=\frac{J_\text{s}^K}{(4I)^{K-1}}$} by proving \mbox{$\sum_{\sigma\in S_K}\prod_{n=1}^{K-1}\frac{1}{\# QP(\sigma,n)}=2^{K-1}$} by induction on $K$. The effective coupling for the bilayer toric code \cite{Wiedmann_2020} \mbox{$J_\text{s}^\text{eff}=J_\text{s}^2/4I$} can be used as the base case. The induction step is equivalent to the statement
	\begin{equation}\label{step}
	\sum_{\sigma\in S_{K+1}}\prod_{n=1}^{K}\frac{1}{\# QP(\sigma,n)}=2\sum_{\sigma\in S_K}\prod_{n=1}^{K-1}\frac{1}{\# QP(\sigma,n)}.
	\end{equation}
	The proof of Eq.~\eqref{step} is simplified by the formula 
	\begin{equation}\label{lemma}
	\begin{split}
	&\frac{1}{a_1a_2\dots a_N}=\\
	&\frac{1}{(a_1+1)a_1a_2\dots a_N}+\frac{1}{(a_1+1)(a_2+1)a_2a_3\dots a_N}+\dots\\
	&+\frac{1}{(a_1+1)(a_2+1)\dots(a_N+1)a_N}\\
	&+\frac{1}{(a_1+1)(a_2+1)\dots(a_N+1)}
	\end{split}
	\end{equation}
	for $a_1,\dots,a_N\in \mathbb{R}\setminus \{-1,0\}$ and $N\in \mathbb{N}$ which can be proven by induction. Now, the induction step for the proof of $J_\text{s}^\text{eff}=\frac{J_\text{s}^K}{(4I)^{K-1}}$ from Eq.~\eqref{step} can be shown. The approach is to split the right hand side of Eq.~\eqref{step} into its addends and to show that the equality holds for every addend individually for a suitable choice of addends of the left hand side. For this, the set $\Omega_\sigma\subset S_{K+1}$ is defined as
	\begin{equation}
	\begin{split}
	&\Omega_\sigma\equiv\Big\{\big(K+1,\sigma(1),\dots,\sigma(K)\big),\\
	&\big(\sigma(1),K+1,\sigma(2)\dots,\sigma(K)\big),\dots,\big(\sigma(1),\dots,\sigma(K),K+1\big)\Big\}.
	\end{split}
	\end{equation}
	For clearity, permutations of $S_K$ are denoted by \mbox{$\sigma=(\sigma(1),\dots,\sigma(K))$} and permutations of $S_{K+1}$ by $\tau$. Furthermore, the set $\Omega_\sigma$ is divided into two disjoint subsets
	\begin{equation}
	\Omega_\sigma^1\equiv\Big\{\tau\in\Omega_\sigma|\tau^{-1}(K+1)<\tau^{-1}(K)\Big\}
	\end{equation}
	and
	\begin{equation}
	\Omega_\sigma^2\equiv\Big\{\tau\in\Omega_\sigma|\tau^{-1}(K+1)>\tau^{-1}(K)\Big\}.
	\end{equation}
	As intended, the disjoint union of all sets $\Omega_\sigma^{1,2}$ 
	\begin{equation}
	\dot\bigcup_{\sigma\in S_K}\Omega_\sigma^1\dot\cup\,\Omega_\sigma^2=S_{K+1}
	\end{equation}
	is the symmetric group $S_{K+1}$. Intuitively, $\Omega_\sigma$ is the set of orders in which the spins $1,\dots,K$ are flipped in the order given by $\sigma$ and the spin $K+1$ is flipped in an arbitrary step in between. $\Omega_\sigma^{1,2}$ distinguish between the cases that spin $K+1$ is flipped before and after its only neighbor $K$. Defining $j$ as $\sigma^{-1}(K)$, it follows that
	\begin{equation}
	K=\begin{cases}
	\tau(j+1)&\tau\in\Omega_\sigma^1\\
	\tau(j)&\tau\in\Omega_\sigma^2
	\end{cases}
	\end{equation}
	which means that spin $K$ is flipped by the $j$-th $V$ in $\Omega_\sigma^2$ and by the $(j+1)$-th $V$ in $\Omega_\sigma^1$.
	
	Now, the contributions from permutations of $\Omega_\sigma^{1,2}$ to the sum on the left hand side of Eq.~\eqref{step} are to be calculated with $p_n\equiv\#QP(\sigma,n)$. For each addend, it is denoted in which step spin $K+1$ is flipped by the corresponding permutation $\tau$. Using this, the modification of the term $(p_1p_2\dots p_{K-1})^{-1}$ from the process given by $\sigma$ can be expressed by adding 1 to $p_n$ in all steps with an additional quasiparticle from the spins $K$ and $K+1$. In this form the contributions from $\Omega_\sigma^{1,2}$ read
	\begin{equation*}
	\begin{split}
	\sum_{\tau\in\Omega_\sigma^1}&\prod_{n=1}^K\frac{1}{\#QP(\tau,n)}=\\
	=&\underbrace{\frac{1}{1\cdot(p_1+1)(p_2+1)\dots(p_{j-1}+1)p_j\dots p_{K-1}}}_{\tau(1)=K+1}\\
	&+\underbrace{\frac{1}{p_1(p_1+1)(p_2+1)\dots(p_{j-1}+1)p_j\dots p_{K-1}}}_{\tau(2)=K+1}\\
	&+\underbrace{\frac{1}{p_1p_2(p_2+1)(p_3+1)\dots(p_{j-1}+1)p_j\dots p_{K-1}}}_{\tau(3)=K+1}\\&+\dots\\
	&+\underbrace{\frac{1}{p_1p_2\dots p_{j-1}(p_{j-1}+1)p_j\dots p_{K-1}}}_{\tau(j)=K+1}=\\
	\stackrel{\text{(\ref{lemma})}}{=}\;&\frac{1}{p_1p_2\dots p_{j-1}}\frac{1}{p_j\dots p_{K-1}}=\prod_{n=1}^{K-1}\frac{1}{\#QP(\sigma,n)}\\
	\end{split}
	\end{equation*}
	and
	\begin{equation*}
	\begin{split}
	&\sum_{\tau\in\Omega_\sigma^2}\prod_{n=1}^K\frac{1}{\#QP(\tau,n)}=\\
	&=\underbrace{\frac{1}{p_1\dots p_{j-1}(p_j+1)p_jp_{j+1}\dots p_{K-1}}}_{\tau(j+1)=K+1}\\
	&+\underbrace{\frac{1}{p_1\dots p_{j-1}(p_j+1)(p_{j+1}+1)p_{j+1}p_{j+2}\dots p_{K-1}}}_{\tau(j+2)=K+1}\\
	&+\underbrace{\frac{1}{p_1\dots p_{j-1}(p_j+1)(p_{j+1}+1)(p_{j+2}+1)p_{j+2}p_{j+3}\dots p_{K-1}}}_{\tau(j+3)=K+1}\\&+\dots\\
	&+\underbrace{\frac{1}{p_1\dots p_{j-1}(p_j+1)(p_{j+1}+1)\dots (p_{K-1}+1)p_{K-1}}}_{\tau(K)=K+1}\\
	&+\underbrace{\frac{1}{p_1\dots p_{j-1}(p_j+1)(p_{j+1}+1)\dots (p_{K-1}+1)\cdot 1}}_{\tau(K+1)=K+1}=\\
	&\stackrel{\text{(\ref{lemma})}}{=}\;\frac{1}{p_1p_2\dots p_{j-1}}\frac{1}{p_j\dots p_{K-1}}=\prod_{n=1}^{K-1}\frac{1}{\#QP(\sigma,n)}.
	\end{split}
	\end{equation*}
	In the first calculation, Eq.~\eqref{lemma} was applied with $N=j-1$, $a_1=p_{j-1}$ and $a_{N}=p_1$, and in the second calculation with $N=K-j$, $a_1=p_j$ and $a_{N}=p_{K-1}$. These results imply Eq.~\eqref{step} directly:
	\begin{equation*}
	\begin{aligned}
	&\sum_{\tau\in S_{K+1}}\prod_{n=1}^{K}\frac{1}{\#QP(\tau,n)}\\
	&=\sum_{\sigma\in S_K}\left(\sum_{\tau\in\Omega_\sigma^1}\prod_{n=1}^{K}\frac{1}{\#QP(\tau,n)}+\sum_{\tau\in\Omega_\sigma^2}\prod_{n=1}^{K}\frac{1}{\#QP(\tau,n)}\right)\\
	&=2\sum_{\sigma\in S_K}\prod_{n=1}^{K-1}\frac{1}{\# QP(\sigma,n)}.
	\end{aligned}
	\end{equation*}
	
%Methods
%%%%%%%%%%%%%%%%%%%%%%%%%%%%%%%%%%%%%%%%%%%%%%%%%%%%%%%%%%%%%%%%%%%%%%%%%%%%%%%%%%%%%%%%%%%%
\section{Technical aspects}
\label{app_technical_aspects}
In the following we give technical aspects on L\"owdin's partition technique, pCUTs, as well as on DLog Pad\'{e} extrapolation.
\subsection{L\"owdin's partition technique}

For an eigenvalue problem
\begin{equation}
\left( \mathcal{H}_0 + \lambda V \right) \ket{\Psi_i} = E_i \ket{\Psi_i}
\end{equation}
L\"owdin's partioning technique \cite{Loewdin1962, Yao2000, Kalis_2012} can be used to determine the eigenvalues $E_i$ up to a given order in $\lambda$. 
To this sake the Hilbert space is decomposed into the eigenspace of interest and its orthogonal complement i.e. a state $\ket{\Psi_i}$ is decomposed into
\begin{equation} \ket{\Psi_i} = P \ket{\Psi_i} + Q \ket{\Psi_i} = \ket{\Psi_i}_p + \ket{\Psi_i}_q \, \end{equation}
where $P$ is a projection operator onto the eigenspace of interest and $Q= 1 - P$ projects onto the orthogonal complement of this subspace \cite{Yao2000}.
After some rearrangement  \cite{Griffin_2008} one can write an eigenvalue equation of the form 
\begin{equation}
\Theta_i \ket{\Psi_i}_p = \left( E_i - E_i^{(0)} \right) \ket{\Psi_i}_p \, ,
\end{equation}
where $E_i$ is the energy eigenvalue of $\mathcal{H}$ and $E^{(0)}_i$ is the energy eigenvalue of $\mathcal{H}_0$ for any state in the eigenspace of interest.
The operator $\Theta_i $ can then be shown to be given as \cite{Kalis_2012}
\begin{equation}
\Theta_i  = PV  \Bigg[ \sum_{m=0}^{\infty} \Bigg( \sum_{j=0}^{\infty}   \Big( -S \sum_{k=1}^{\infty} E_i^{(k)} \Big)^j SV \Bigg)^m  \Bigg] P \, ,
\end{equation}
where $E_i^{(k)}$ is the energy in order $k$ and
\begin{equation}
S = \frac{Q}{E^{(0)}_i - \mathcal{H}_0} \, .
\end{equation}
Note that in order to calculate $\Theta_i$ up to order $k$ only energy corrections upto order $k-2$ are needed during the calculation, so this operator can be calculated iteratively upto a desired order in $V$.
Actually the iterative use of the energy corrections results efficient, not only because these terms do not have to be recalculated, but also because it is easy to sort out terms based on very simple criteria. For example, if no first-order correction exists, all terms involving the corresponding energy can be dropped \cite{Kalis_2012}.
Obviously, expectation values calculated with Löwdin's partition technique equal those calculated with pCUT. 
Accordingly, it is usually suitable to perform the linked cluster expansions for ground-state energies with L\"owdin instead of pCUT. \par

\subsection{Method: pCUT}

For perturbative continuous unitary transformations (pCUTs) \cite{Knetter_2000, Knetter2003} we start with a lattice Hamiltonian, which has an equidistant spectrum and is bounded from below 
\begin{equation} \mathcal{H} = E_0 + Q + \lambda \sum_{n = -N}^N T_n \,  ,\end{equation} 
where $Q$ is a quasiparticle-counting operator and the $T_n$-operators change the number of quasi-particles (QPs) by $n$
\begin{equation}
[Q, T_n] = n T_n \, .
\end{equation}
The pCUT method maps such a many-particle lattice Hamiltonian
to a QP-conserving effective Hamiltonian \cite{Knetter_2000, Knetter2003}
\begin{equation} \mathcal{H}_{\text{eff}} = E_0 + Q + \sum_k \lambda^k  \sum_{ \substack{\vec{m}, \vert \vec{m} \vert = k \\ \sum_i m_i = 0} }   \mathcal{C}(\vec{m})  T(\vec{m}) \,   \end{equation}
with
\begin{align}
\vec{m} &= (m_1, \ldots, m_k) \, ,\\
\vert \vec{m} \vert &= \text{dim}(\vec{m}) \, ,\\
T(\vec{m}) &= T_{m_1} \ldots T_{m_k} \, . 
\end{align}
Interestingly the effective Hamiltonian can be rewritten as the sum of nested commutators of $T$-operators \cite{Dusuel2010}.
This naturally leads to a linked cluster property, i.e. that all processes involved in the perturbative expansion are defined on connected subclusters.
This makes the method naturally suited for linked cluster expansions \cite{coester2015optimizing} in all quasi-particle sectors, which we also employ in this paper, in order to push to higher perturbation orders.

%Methods
%%%%%%%%%%%%%%%%%%%%%%%%%%%%%%%%%%%%%%%%%%%%%%%%%%%%%%%%%%%%%%%%%%%%%%%%%%%%%%%%%%%%%%%%%%%%
\subsection{Method: DLog Pad\'{e} approximation}

%DLog Padé approdimation
%%%%%%%%%%%%%%%%%%%%%%%KITC%%%%%%%%%%%%%%%%%%%%%%%%%%%%%%%%%%%%%%%%%%%%%%%%%%%%%%%%%%%%%%%%%%%%%
For a more precise analysis of the system regarding phase transitions, the ground-state energies and energy gaps in both limiting cases must be investigated for parameter values of $I/J_\text{s}$ in the vicinity of the points the gaps close at. Here the ground-state energy and the energy gap in the low-Ising limit are denoted by $E_{0,\rm lI}$ and $\Delta_{\rm lI}$ and in the high-Ising limit by $E_{0,\rm hI}$ and $\Delta_{\rm hI}$. Besides the zero points of the gaps corresponding to phase transition points, also the dependence of these quantities on the parameter controlling the transition is interesting since, in the case of a second-order phase transitions, close to the critical point these dependencies are described by critical exponents which are determined by the universality class of the phase transition. In general, with $x$ being a normalized model parameter, the energy gap scales according to
\begin{equation}
	\Delta\propto(x-x_c)^{\nu z}\text{ for }|x-x_c|\ll1
\end{equation}
with the critical exponents $\nu$ describing the correlation length and the dynamical critical exponent $z$ describing the autocorrelation time. The second derivative of the ground-state energy diverges at the critical point which is described by the exponent $\alpha$
\begin{equation}
	\frac{d^2 E_0}{d x^2}\propto(x-x_c)^{-\alpha}\text{ for }|x-x_c|\ll1,
\end{equation}
corresponding to the behavior of the specific heat in thermal phase transitions. In order to gain access to these quantities higher orders of the perturbative series are needed. Therefore, an implementation for computing these corrections was used to obtain the exact series expansions of the ground-state energies up to order 14, the low-Ising gap up to order 11, and the high-Ising gap up to order 8. The ground-state energies were evaluated using a full graph decomposition and a perturbative method by L\"owdin \cite{Loewdin1962}, the gap in the low-Ising limit using perturbative continuous unitary transformations \cite{Knetter_2000} and graph decomposition, and the gap in the high-Ising limit on a cluster using the Takahashi method \cite{Takahashi_1977,Klagges_2012}. The results are given in App.~\ref{app_series}. For the investigation of quantum critical behavior however, the convergence of the plain series expansions is not yet expected to be satisfying since at the critical point $x_c$ the observables depend algebraically on the parameter $x$.

The approximations can be refined by performing Pad\'{e} extrapolations. Instead of polynomials, rational functions are used with the same condition that all known derivatives must coincide with the expansion. These rational approximations are much more suitable to describe functions close to poles than the raw series expansions. In order to apply this technique on observables with an undifferentiable but not diverging point, the observable needs to be modified. Considering the energy gap again, differentiating the logarithm of the gap
\begin{equation}\label{PadeEx}
	\frac{d}{dx}\log{\Delta}=\frac{\Delta'}{\Delta}\sim\frac{\nu z}{x-x_c}
\end{equation}
yields the desired form. In principle, a pole structure could also be obtained by differentiating the gap only, but the logarithm gives better numerical access to the exponent, then equaling the residue
\begin{equation}
	\nu z=\left[(x-x_c)\frac{d}{dx}\log{\Delta}\right]_{x_c}.
\end{equation}
Pad\'{e} approximations of such modified functions $f$ are also called DLog Pad\'{e} approximations and defined in terms of the chosen degrees of the polynomials in the enumerator $m$ and in the denominator $n$ as
\begin{align}
	\text{DLog Pad\'{e}}[m,n](f)=\frac{q(x)}{p(x)}=\frac{\sum_{i=0}^ma_ix^i}{1+\sum_{i=1}^nb_ix^i}\approx\frac{d}{dx}\log{f}.
\end{align}
The coefficient $b_0$ can be set to one for $p(0)\neq 0$ by rescaling all other coefficients. In this form all $m+n+1$ coefficients are determined by the $N+1$ coefficients obtained from a series expansion of $\frac{d}{dx}\log{f}$ up to order $N$ if $m+n=N$ holds, using the condition that the derivatives coincide. If instead the series expansion of $f$ up to order $N$ is used, the highest order is distorted by the derivation and the condition for the highest correct approximation reads $m+n=N-1$.

However, the quality of the individual approximants for degrees $m$ and $n$ depends on how suitable the corresponding rational function is to the physical behavior. In order to avoid artifacts in the extrapolation, approximants possessing certain properties will be neglected. Firstly, if an approximant contains additional poles close to the assumingly physical pole, the physical pole must be expected to be displaced. Also complex poles must be considered disturbing. For the later discussion, all approximants with additional poles within the radius of $I/J_\text{s}=0.2$ around the critical point will be neglected, therefore. Secondly, if $q$ and $p$ share roots, the resulting values equal the ones from the approximant DLog Pad\'{e}$[m-1,n-1]$ which then are only included in the analysis for the lower order. Neglected poles are also referred to as defective. We do not expect approximants with $m=1$ or $n=1$ to successfully indicate the critical values and thus neglect them.
\FloatBarrier
\onecolumngrid

%Series (Ground state energy and gap for K=3 and K=4)
%%%%%%%%%%%%%%%%%%%%%%%%%%%%%%%%%%%%%%%%%%%%%%%%%%%%%%%%%%%%%%%%%%%%%%%%%%%%%%%%%%%%%%%%%%%%
\section{High-order series for $K=3$ and $K=4$}
\label{app_series}

\setlength{\tabcolsep}{6pt} % General space between cols (6pt standard)
\renewcommand{\arraystretch}{1.4} % General space between rows (1 standard)
\begin{table*}[h!]
    \centering
    \begin{tabular}{r||r|r}
    $n$& $\Delta_{\rm lI}$ &$\Delta_{\rm hI}$\\ \hline
     0& $4$ &  $8$\\
     1& $-4$ &  $-1$\\
     2& $-\frac{3}{2}$& $-\frac{37}{48}$\\
     3& $-\frac{9}{16}$& $-\frac{11}{48}$\\
     4& $-\frac{87}{64}$& $-\frac{42127}{552960}$\\
     5& $-\frac{1079}{3072}$& $\frac{188473}{11059200}$\\
     6& $-\frac{18303}{8192}$& $\frac{761018983}{222953472000}$\\
     7& $-\frac{296207}{10616832}$& $-\frac{184224828979}{10404495360000}$\\
     8& $-\frac{4784639425}{1019215872}$&$-\frac{869230920151914301} { 83062832077209600000}$\\
     9& $\frac{67928594161}{50960793600}$&\\
     10& $-\frac{1660420331076899}{146767085568000}$&\\
     11& $\frac{225188034597766031}{35224100536320000}$&
\end{tabular}
    \caption{Coefficients of the expansions of the energy gaps in the low-Ising and the high-Ising limit of the 3ITC. A coefficient of order $n$ contributes to the expansion proportional to $J_\text{s}^n/I^{n-1}$ in the high-Ising limit and to $I^n/J_\text{s}^{n-1}$ in the low-Ising limit.}
    \label{GapLIK3}
    \label{GapHIK3}
\end{table*}

\begin{table*}
    \centering
    \begin{tabular}{r||r|r}
    $n$& $\epsilon_{0,\rm lI}$ & $\epsilon_{0,\rm hI}$\\ \hline
     0& $-3$&$-4$  \\
     1& $0$ & 0\\
     2& $-\frac{1}{2}$&$-\frac{5}{16}$ \\
     3& 0&$-\frac{1}{16}$   \\
     4& $-\frac{9}{64}$&$-\frac{589}{61440}$  \\
     5& 0&$-\frac{131}{51200}$  \\
     6& $-\frac{48505}{442368}$&$-\frac{9126259}{8257536000}$  \\
     7& 0&$-\frac{617369659}{1541406720000}$  \\
     8& $-\frac{330744647}{2548039680}$&$-\frac{1310971599346159}{9229203564134400000}$  \\
     9& 0&$-\frac{1766730475174727261}{28425946977533952000000}$  \\
     10& $-\frac{9187858198193}{48922361856000}$&$-\frac{43704772621217954597025937}{1475074692406675431751680000000}$ \\
     11& 0&$-\frac{14749740844849694201490005304457}{1063115832311339117172070809600000000}$ \\
     12& $-\frac{3037172238388823419}{9862748150169600000}$&$-\frac{246780583842162456001212650178268543}{37837473711774238445839746859008000000000}$ \\
     13& 0&$-\frac{66159875509906659470470524127089907055747}{21239309118630233266803225104366960640000000000}$ \\
     14& $-\frac{91769629940536515750103099}{167019722274232074240000000}$&$-\frac{23312415467805500378675766210538051106685250654307}{15465044415192681874885224721305040675720396800000000000}$ 
    \end{tabular}
    \caption{Coefficients of the low-Ising and high-Ising ground-state energy expansions per unit cell for the 3ITC. A coefficient of order $n$ contributes to the expansion proportional to $J_\text{s}^{n}/I^{n-1}$ in the high-Ising limit and to $I^{n}/J_\text{s}^{n-1}$ in the low-Ising limit. The energy zero point of the zeroth order is chosen according to the model in its original form of Eq.~\eqref{HKITC}.}
    \label{GSLIK3}
    \label{GSHIK3}
\end{table*}

\begin{table*}[]
    \centering
    \begin{tabular}{r||r|r}
    $n$& $\Delta_{\rm lI}$ &$\Delta_{\rm hI}$ \\ \hline
     0& $4$& $8$\\
     1& $-4$& $-\sqrt{2}$\\
     2& $-2$&	$-\frac{7}{16}$\\
     3& $-\frac{3}{8}$& $\frac{355\sqrt{2}}{9216}$\\
     4& $-\frac{69}{32}$& $-\frac{86983}{483840}$\\
     5& $-\frac{73}{3072}$& $\frac{682663741\sqrt{2}}{41617981440}$\\
     6& $-\frac{5969}{1536}$& $-\frac{163885370221}{10924720128000}$\\
     7& $\frac{15834157}{21233664 }$&$\frac{4657908195293447\sqrt{2}}{4228653269385216000}$\\
     8& $-\frac{22304396311}{2548039680}$&\\
     9& $\frac{2171225541259}{611529523200}$&\\
     10& $-\frac{3231402801984019}{146767085568000}$&\\
     11& $\frac{911179726263954809}{70448201072640000}$&
\end{tabular}
    \caption{Coefficients of the expansions of the energy gaps in the low-Ising and the high-Ising limit of the 4ITC. A coefficient of order $n$ contributes to the expansion proportional to $J_\text{s}^n/I^{n-1}$ in the high-Ising limit and to $I^n/J_\text{s}^{n-1}$ in the low-Ising limit.}
    \label{GapK4}
\end{table*}

\begin{table*}
    \centering
    \begin{tabular}{r||r|r}
    $n$& $\epsilon_{0,\rm lI}$ & $\epsilon_{0,\rm hI}$\\ \hline
     0& $-4$&$-6$ \\
     1& $0$ & 0\\
     2& $-\frac{3}{4}$&$-\frac{3}{8}$ \\
     3& 0 & 0 \\
     4& $-\frac{57}{256}$&$-\frac{467}{17920}$  \\
     5& 0 & 0\\
     6& $-\frac{41875}{221184}$&$-\frac{525334909}{303464448000}$  \\
     7& 0 & 0\\
     8& $-\frac{83376133}{339738624}$&$-\frac{1265026645996681769}{4409252002765209600000}$  \\
     9& 0 & 0\\
     10& $-\frac{22902916849889}{58706834227200}$&$-\frac{1272424614372127280829605158291}{27483960437594278313255239680000000}$  \\
     11& 0 & 0\\
     12& $-\frac{83282317722082165181}{118352977802035200000}$&$-\frac{41341338783438993572322990193714218652576621}{4611210890915158979173246330343497138176000000000}$  \\
     13& 0 & 0\\
     14& $-\frac{17101800787552558640004551}{12371831279572746240000000}$&$-\frac{5106623631618246901094489839619918742305524098826776618502037}{2665519726676254993620521639222095455960274480218400358400000000000}$ 
    \end{tabular}
    \caption{Coefficients of the low-Ising and high-Ising ground-state energy expansions per  unit cell for the 4ITC. A coefficient of order $n$ contributes to the expansion proportional to $J_\text{s}^{n}/I^{n-1}$ in the high-Ising limit and to $I^{n}/J_\text{s}^{n-1}$ in the low-Ising limit. The energy zero point of the zeroth order is chosen according to the model in its original form of Eq.~\eqref{HKITC}.}
    \label{GSK4}
\end{table*}

%Results K=4
%%%%%%%%%%%%%%%%%%%%%%%%%%%%%%%%%%%%%%%%%%%%%%%%%%%%%%%%%%%%%%%%%%%%%%%%%%%%%%%%%%%%%%%%%%%%
%\section{Results $K=4$}	

%\renewcommand{\arraystretch}{1.2}
%\setlength{\tabcolsep}{12pt} % General space between cols (6pt standard)
%\begin{table}[tb]
%	\centering
%	\begin{tabular}{l|ll}
%											&lI-limit 	&hI-limit \\\hline
%		Energy gap 							&$0.5729(5)$&    \\
%		Ground-state energy  				&$0.581(2)$ & $0.55(1)$
%		
%	\end{tabular}
%	\caption{Critical points with sample standard deviations of all reliable expansions for $K=4$. Highest approximants of families with more than one member form the samples apart from the gap in the lI-limit, where also single-approximant families are included. The available estimate for the exponent from the gap in the lI-limit $\nu z$ is $0.567(6)$. \leacomment{Das ist die gleiche Tabelle wie II. Brauchen wir die zweimal?}}
%	\label{tabel}
%\end{table}

%\vfill
\end{appendix}

\FloatBarrier % Prevent images to slip into Refs
\twocolumngrid
\newpage


\begin{thebibliography}{66}%
\makeatletter
\providecommand \@ifxundefined [1]{%
 \@ifx{#1\undefined}
}%
\providecommand \@ifnum [1]{%
 \ifnum #1\expandafter \@firstoftwo
 \else \expandafter \@secondoftwo
 \fi
}%
\providecommand \@ifx [1]{%
 \ifx #1\expandafter \@firstoftwo
 \else \expandafter \@secondoftwo
 \fi
}%
\providecommand \natexlab [1]{#1}%
\providecommand \enquote  [1]{``#1''}%
\providecommand \bibnamefont  [1]{#1}%
\providecommand \bibfnamefont [1]{#1}%
\providecommand \citenamefont [1]{#1}%
\providecommand \href@noop [0]{\@secondoftwo}%
\providecommand \href [0]{\begingroup \@sanitize@url \@href}%
\providecommand \@href[1]{\@@startlink{#1}\@@href}%
\providecommand \@@href[1]{\endgroup#1\@@endlink}%
\providecommand \@sanitize@url [0]{\catcode `\\12\catcode `\$12\catcode
  `\&12\catcode `\#12\catcode `\^12\catcode `\_12\catcode `\%12\relax}%
\providecommand \@@startlink[1]{}%
\providecommand \@@endlink[0]{}%
\providecommand \url  [0]{\begingroup\@sanitize@url \@url }%
\providecommand \@url [1]{\endgroup\@href {#1}{\urlprefix }}%
\providecommand \urlprefix  [0]{URL }%
\providecommand \Eprint [0]{\href }%
\providecommand \doibase [0]{https://doi.org/}%
\providecommand \selectlanguage [0]{\@gobble}%
\providecommand \bibinfo  [0]{\@secondoftwo}%
\providecommand \bibfield  [0]{\@secondoftwo}%
\providecommand \translation [1]{[#1]}%
\providecommand \BibitemOpen [0]{}%
\providecommand \bibitemStop [0]{}%
\providecommand \bibitemNoStop [0]{.\EOS\space}%
\providecommand \EOS [0]{\spacefactor3000\relax}%
\providecommand \BibitemShut  [1]{\csname bibitem#1\endcsname}%
\let\auto@bib@innerbib\@empty
%</preamble>
\bibitem [{\citenamefont {Wen}(1989)}]{Wen_1989}%
  \BibitemOpen
  \bibfield  {author} {\bibinfo {author} {\bibfnamefont {X.-G.}\ \bibnamefont
  {Wen}},\ }\href {https://doi.org/10.1103/PhysRevB.40.7387} {\bibfield
  {journal} {\bibinfo  {journal} {Phys. Rev. B}\ }\textbf {\bibinfo {volume}
  {40}},\ \bibinfo {pages} {7387} (\bibinfo {year} {1989})}\BibitemShut
  {NoStop}%
\bibitem [{\citenamefont {Wen}(1990)}]{Wen_1990}%
  \BibitemOpen
  \bibfield  {author} {\bibinfo {author} {\bibfnamefont {X.-G.}\ \bibnamefont
  {Wen}},\ }\href {https://doi.org/10.1142/S0217979290000139} {\bibfield
  {journal} {\bibinfo  {journal} {Int. J. Mod. Phys. B}\ }\textbf {\bibinfo
  {volume} {4}},\ \bibinfo {pages} {239} (\bibinfo {year} {1990})}\BibitemShut
  {NoStop}%
\bibitem [{\citenamefont {Wen}(2004)}]{Wen_2004}%
  \BibitemOpen
  \bibfield  {author} {\bibinfo {author} {\bibfnamefont {X.-G.}\ \bibnamefont
  {Wen}},\ }\href@noop {} {\emph {\bibinfo {title} {Quantum Field Theory of
  Many-body Systems: From the Origin of Sound to an Origin of Light and
  Electrons}}}\ (\bibinfo  {publisher} {Oxford University Press},\ \bibinfo
  {year} {2004})\BibitemShut {NoStop}%
\bibitem [{\citenamefont {Leinaas}\ and\ \citenamefont
  {Myrheim}(1977)}]{Leinaas_1977}%
  \BibitemOpen
  \bibfield  {author} {\bibinfo {author} {\bibfnamefont {J.~M.}\ \bibnamefont
  {Leinaas}}\ and\ \bibinfo {author} {\bibfnamefont {J.}~\bibnamefont
  {Myrheim}},\ }\href {https://doi.org/10.1007/BF02727953} {\bibfield
  {journal} {\bibinfo  {journal} {Il Nuovo Cimento B}\ }\textbf {\bibinfo
  {volume} {37}},\ \bibinfo {pages} {1} (\bibinfo {year} {1977})}\BibitemShut
  {NoStop}%
\bibitem [{\citenamefont {Wilczek}(1982)}]{Wilczek_1982}%
  \BibitemOpen
  \bibfield  {author} {\bibinfo {author} {\bibfnamefont {F.}~\bibnamefont
  {Wilczek}},\ }\href {https://doi.org/10.1103/PhysRevLett.48.1144} {\bibfield
  {journal} {\bibinfo  {journal} {Phys. Rev. Lett.}\ }\textbf {\bibinfo
  {volume} {48}},\ \bibinfo {pages} {1144} (\bibinfo {year}
  {1982})}\BibitemShut {NoStop}%
\bibitem [{\citenamefont {Kitaev}\ and\ \citenamefont
  {Preskill}(2006)}]{Kitaev_2006_b}%
  \BibitemOpen
  \bibfield  {author} {\bibinfo {author} {\bibfnamefont {A.}~\bibnamefont
  {Kitaev}}\ and\ \bibinfo {author} {\bibfnamefont {J.}~\bibnamefont
  {Preskill}},\ }\href {https://doi.org/10.1103/PhysRevLett.96.110404}
  {\bibfield  {journal} {\bibinfo  {journal} {Phys. Rev. Lett.}\ }\textbf
  {\bibinfo {volume} {96}},\ \bibinfo {pages} {110404} (\bibinfo {year}
  {2006})}\BibitemShut {NoStop}%
\bibitem [{\citenamefont {Levin}\ and\ \citenamefont {Wen}(2006)}]{Levin_2006}%
  \BibitemOpen
  \bibfield  {author} {\bibinfo {author} {\bibfnamefont {M.~A.}\ \bibnamefont
  {Levin}}\ and\ \bibinfo {author} {\bibfnamefont {X.-G.}\ \bibnamefont
  {Wen}},\ }\href {https://doi.org/10.1103/PhysRevLett.96.110405} {\bibfield
  {journal} {\bibinfo  {journal} {Phys. Rev. Lett.}\ }\textbf {\bibinfo
  {volume} {96}},\ \bibinfo {pages} {110405} (\bibinfo {year}
  {2006})}\BibitemShut {NoStop}%
\bibitem [{\citenamefont {Kitaev}(2003)}]{Kitaev_2003}%
  \BibitemOpen
  \bibfield  {author} {\bibinfo {author} {\bibfnamefont {A.}~\bibnamefont
  {Kitaev}},\ }\href {https://doi.org/10.1016/S0003-4916(02)00018-0} {\bibfield
   {journal} {\bibinfo  {journal} {Ann. Phys.}\ }\textbf {\bibinfo {volume}
  {303}},\ \bibinfo {pages} {2} (\bibinfo {year} {2003})}\BibitemShut {NoStop}%
\bibitem [{\citenamefont {Nayak}\ \emph {et~al.}(2008)\citenamefont {Nayak},
  \citenamefont {Simon}, \citenamefont {Stern}, \citenamefont {Freedman},\ and\
  \citenamefont {Sarma}}]{Nayak_2008}%
  \BibitemOpen
  \bibfield  {author} {\bibinfo {author} {\bibfnamefont {C.}~\bibnamefont
  {Nayak}}, \bibinfo {author} {\bibfnamefont {S.~H.}\ \bibnamefont {Simon}},
  \bibinfo {author} {\bibfnamefont {A.}~\bibnamefont {Stern}}, \bibinfo
  {author} {\bibfnamefont {M.}~\bibnamefont {Freedman}},\ and\ \bibinfo
  {author} {\bibfnamefont {S.~D.}\ \bibnamefont {Sarma}},\ }\href
  {https://doi.org/10.1103/RevModPhys.80.1083} {\bibfield  {journal} {\bibinfo
  {journal} {Rev. Mod. Phys.}\ }\textbf {\bibinfo {volume} {80}},\ \bibinfo
  {pages} {1083} (\bibinfo {year} {2008})}\BibitemShut {NoStop}%
\bibitem [{\citenamefont {Hamma}\ \emph {et~al.}(2005)\citenamefont {Hamma},
  \citenamefont {Zanardi},\ and\ \citenamefont {Wen}}]{Hamma_2005}%
  \BibitemOpen
  \bibfield  {author} {\bibinfo {author} {\bibfnamefont {A.}~\bibnamefont
  {Hamma}}, \bibinfo {author} {\bibfnamefont {P.}~\bibnamefont {Zanardi}},\
  and\ \bibinfo {author} {\bibfnamefont {X.-G.}\ \bibnamefont {Wen}},\ }\href
  {https://doi.org/10.1103/PhysRevB.72.035307} {\bibfield  {journal} {\bibinfo
  {journal} {Phys. Rev. B}\ }\textbf {\bibinfo {volume} {72}},\ \bibinfo
  {pages} {035307} (\bibinfo {year} {2005})}\BibitemShut {NoStop}%
\bibitem [{\citenamefont {Nussinov}\ and\ \citenamefont
  {Ortiz}(2008)}]{Nussinov_2008}%
  \BibitemOpen
  \bibfield  {author} {\bibinfo {author} {\bibfnamefont {Z.}~\bibnamefont
  {Nussinov}}\ and\ \bibinfo {author} {\bibfnamefont {G.}~\bibnamefont
  {Ortiz}},\ }\href {https://doi.org/10.1103/PhysRevB.77.064302} {\bibfield
  {journal} {\bibinfo  {journal} {Phys. Rev. B}\ }\textbf {\bibinfo {volume}
  {77}},\ \bibinfo {pages} {064302} (\bibinfo {year} {2008})}\BibitemShut
  {NoStop}%
\bibitem [{\citenamefont {Reiss}\ and\ \citenamefont
  {Schmidt}(2019)}]{Reiss_2019}%
  \BibitemOpen
  \bibfield  {author} {\bibinfo {author} {\bibfnamefont {D.~A.}\ \bibnamefont
  {Reiss}}\ and\ \bibinfo {author} {\bibfnamefont {K.~P.}\ \bibnamefont
  {Schmidt}},\ }\href {https://doi.org/10.21468/SciPostPhys.6.6.078} {\bibfield
   {journal} {\bibinfo  {journal} {SciPost Phys.}\ }\textbf {\bibinfo {volume}
  {6}},\ \bibinfo {pages} {78} (\bibinfo {year} {2019})}\BibitemShut {NoStop}%
\bibitem [{\citenamefont {Chamon}(2005)}]{Chamon_2005}%
  \BibitemOpen
  \bibfield  {author} {\bibinfo {author} {\bibfnamefont {C.}~\bibnamefont
  {Chamon}},\ }\href {https://doi.org/10.1103/PhysRevLett.94.040402} {\bibfield
   {journal} {\bibinfo  {journal} {Phys. Rev. Lett.}\ }\textbf {\bibinfo
  {volume} {94}},\ \bibinfo {pages} {040402} (\bibinfo {year}
  {2005})}\BibitemShut {NoStop}%
\bibitem [{\citenamefont {Bravyi}\ \emph {et~al.}(2011)\citenamefont {Bravyi},
  \citenamefont {Leemhuis},\ and\ \citenamefont {Terhal}}]{Bravyi_2011}%
  \BibitemOpen
  \bibfield  {author} {\bibinfo {author} {\bibfnamefont {S.}~\bibnamefont
  {Bravyi}}, \bibinfo {author} {\bibfnamefont {B.}~\bibnamefont {Leemhuis}},\
  and\ \bibinfo {author} {\bibfnamefont {B.~M.}\ \bibnamefont {Terhal}},\
  }\href {https://doi.org/10.1016/j.aop.2010.11.002} {\bibfield  {journal}
  {\bibinfo  {journal} {Ann. Phys.}\ }\textbf {\bibinfo {volume} {326}},\
  \bibinfo {pages} {839} (\bibinfo {year} {2011})}\BibitemShut {NoStop}%
\bibitem [{\citenamefont {Haah}(2011)}]{Haah_2011}%
  \BibitemOpen
  \bibfield  {author} {\bibinfo {author} {\bibfnamefont {J.}~\bibnamefont
  {Haah}},\ }\href {https://doi.org/10.1103/PhysRevA.83.042330} {\bibfield
  {journal} {\bibinfo  {journal} {Phys. Rev. A}\ }\textbf {\bibinfo {volume}
  {83}},\ \bibinfo {pages} {042330} (\bibinfo {year} {2011})}\BibitemShut
  {NoStop}%
\bibitem [{\citenamefont {Yoshida}(2013)}]{Yoshida_2013}%
  \BibitemOpen
  \bibfield  {author} {\bibinfo {author} {\bibfnamefont {B.}~\bibnamefont
  {Yoshida}},\ }\href {https://doi.org/10.1103/PhysRevB.88.125122} {\bibfield
  {journal} {\bibinfo  {journal} {Phys. Rev. B}\ }\textbf {\bibinfo {volume}
  {88}},\ \bibinfo {pages} {125122} (\bibinfo {year} {2013})}\BibitemShut
  {NoStop}%
\bibitem [{\citenamefont {Vijay}\ \emph {et~al.}(2015)\citenamefont {Vijay},
  \citenamefont {Haah},\ and\ \citenamefont {Fu}}]{Vijay_2015}%
  \BibitemOpen
  \bibfield  {author} {\bibinfo {author} {\bibfnamefont {S.}~\bibnamefont
  {Vijay}}, \bibinfo {author} {\bibfnamefont {J.}~\bibnamefont {Haah}},\ and\
  \bibinfo {author} {\bibfnamefont {L.}~\bibnamefont {Fu}},\ }\href
  {https://doi.org/10.1103/PhysRevB.92.235136} {\bibfield  {journal} {\bibinfo
  {journal} {Phys. Rev. B}\ }\textbf {\bibinfo {volume} {92}},\ \bibinfo
  {pages} {235136} (\bibinfo {year} {2015})}\BibitemShut {NoStop}%
\bibitem [{\citenamefont {Vijay}\ \emph {et~al.}(2016)\citenamefont {Vijay},
  \citenamefont {Haah},\ and\ \citenamefont {Fu}}]{Vijay_2016}%
  \BibitemOpen
  \bibfield  {author} {\bibinfo {author} {\bibfnamefont {S.}~\bibnamefont
  {Vijay}}, \bibinfo {author} {\bibfnamefont {J.}~\bibnamefont {Haah}},\ and\
  \bibinfo {author} {\bibfnamefont {L.}~\bibnamefont {Fu}},\ }\href
  {https://doi.org/10.1103/PhysRevB.94.235157} {\bibfield  {journal} {\bibinfo
  {journal} {Phys. Rev. B}\ }\textbf {\bibinfo {volume} {94}},\ \bibinfo
  {pages} {235157} (\bibinfo {year} {2016})}\BibitemShut {NoStop}%
\bibitem [{\citenamefont {M\"uhlhauser}\ \emph {et~al.}(2020)\citenamefont
  {M\"uhlhauser}, \citenamefont {Walther}, \citenamefont {Reiss},\ and\
  \citenamefont {Schmidt}}]{Muehlhauser_2020}%
  \BibitemOpen
  \bibfield  {author} {\bibinfo {author} {\bibfnamefont {M.}~\bibnamefont
  {M\"uhlhauser}}, \bibinfo {author} {\bibfnamefont {M.~R.}\ \bibnamefont
  {Walther}}, \bibinfo {author} {\bibfnamefont {D.~A.}\ \bibnamefont {Reiss}},\
  and\ \bibinfo {author} {\bibfnamefont {K.~P.}\ \bibnamefont {Schmidt}},\
  }\href {https://doi.org/10.1103/PhysRevB.101.054426} {\bibfield  {journal}
  {\bibinfo  {journal} {Phys. Rev. B}\ }\textbf {\bibinfo {volume} {101}},\
  \bibinfo {pages} {054426} (\bibinfo {year} {2020})}\BibitemShut {NoStop}%
\bibitem [{\citenamefont {Mühlhauser}\ \emph {et~al.}(2021)\citenamefont
  {Mühlhauser}, \citenamefont {Schmidt}, \citenamefont {Vidal},\ and\
  \citenamefont {Walther}}]{Muehlhauser_2021}%
  \BibitemOpen
  \bibfield  {author} {\bibinfo {author} {\bibfnamefont {M.}~\bibnamefont
  {Mühlhauser}}, \bibinfo {author} {\bibfnamefont {K.~P.}\ \bibnamefont
  {Schmidt}}, \bibinfo {author} {\bibfnamefont {J.}~\bibnamefont {Vidal}},\
  and\ \bibinfo {author} {\bibfnamefont {M.~R.}\ \bibnamefont {Walther}},\
  }\href@noop {} {\bibinfo {title} {Competing topological orders in three
  dimensions}} (\bibinfo {year} {2021}),\ \Eprint
  {https://arxiv.org/abs/2106.05749} {arXiv:2106.05749 [cond-mat.str-el]}
  \BibitemShut {NoStop}%
\bibitem [{\citenamefont {Ma}\ \emph {et~al.}(2017)\citenamefont {Ma},
  \citenamefont {Lake}, \citenamefont {Chen},\ and\ \citenamefont
  {Hermele}}]{Ma_2017}%
  \BibitemOpen
  \bibfield  {author} {\bibinfo {author} {\bibfnamefont {H.}~\bibnamefont
  {Ma}}, \bibinfo {author} {\bibfnamefont {E.}~\bibnamefont {Lake}}, \bibinfo
  {author} {\bibfnamefont {X.}~\bibnamefont {Chen}},\ and\ \bibinfo {author}
  {\bibfnamefont {M.}~\bibnamefont {Hermele}},\ }\href@noop {} {\bibfield
  {journal} {\bibinfo  {journal} {Phys. Rev. B}\ }\textbf {\bibinfo {volume}
  {95}} (\bibinfo {year} {2017})}\BibitemShut {NoStop}%
\bibitem [{\citenamefont {Vijay}(2017)}]{Vijay_2017}%
  \BibitemOpen
  \bibfield  {author} {\bibinfo {author} {\bibfnamefont {S.}~\bibnamefont
  {Vijay}},\ }\href@noop {} {\bibinfo {title} {Isotropic layer construction and
  phase diagram for fracton topological phases}} (\bibinfo {year} {2017}),\
  \Eprint {https://arxiv.org/abs/1701.00762} {arXiv:1701.00762
  [cond-mat.str-el]} \BibitemShut {NoStop}%
\bibitem [{\citenamefont {Bais}\ \emph {et~al.}(2002)\citenamefont {Bais},
  \citenamefont {Schroers},\ and\ \citenamefont {Slingerland}}]{Bais_2002}%
  \BibitemOpen
  \bibfield  {author} {\bibinfo {author} {\bibfnamefont {F.~A.}\ \bibnamefont
  {Bais}}, \bibinfo {author} {\bibfnamefont {B.~J.}\ \bibnamefont {Schroers}},\
  and\ \bibinfo {author} {\bibfnamefont {J.~K.}\ \bibnamefont {Slingerland}},\
  }\href {https://doi.org/10.1103/PhysRevLett.89.181601} {\bibfield  {journal}
  {\bibinfo  {journal} {Phys. Rev. Lett.}\ }\textbf {\bibinfo {volume} {89}},\
  \bibinfo {pages} {181601} (\bibinfo {year} {2002})}\BibitemShut {NoStop}%
\bibitem [{\citenamefont {Bais}\ and\ \citenamefont {Mathy}(2007)}]{Bais_2007}%
  \BibitemOpen
  \bibfield  {author} {\bibinfo {author} {\bibfnamefont {F.}~\bibnamefont
  {Bais}}\ and\ \bibinfo {author} {\bibfnamefont {C.}~\bibnamefont {Mathy}},\
  }\href {https://doi.org/https://doi.org/10.1016/j.aop.2006.05.010} {\bibfield
   {journal} {\bibinfo  {journal} {Annals of Physics}\ }\textbf {\bibinfo
  {volume} {322}},\ \bibinfo {pages} {552 } (\bibinfo {year}
  {2007})}\BibitemShut {NoStop}%
\bibitem [{\citenamefont {Bais}\ and\ \citenamefont
  {Slingerland}(2009)}]{Bais_2009}%
  \BibitemOpen
  \bibfield  {author} {\bibinfo {author} {\bibfnamefont {F.~A.}\ \bibnamefont
  {Bais}}\ and\ \bibinfo {author} {\bibfnamefont {J.~K.}\ \bibnamefont
  {Slingerland}},\ }\href {https://doi.org/10.1103/PhysRevB.79.045316}
  {\bibfield  {journal} {\bibinfo  {journal} {Phys. Rev. B}\ }\textbf {\bibinfo
  {volume} {79}},\ \bibinfo {pages} {045316} (\bibinfo {year}
  {2009})}\BibitemShut {NoStop}%
\bibitem [{\citenamefont {Burnell}\ \emph {et~al.}(2011)\citenamefont
  {Burnell}, \citenamefont {Simon},\ and\ \citenamefont
  {Slingerland}}]{Burnell_2011}%
  \BibitemOpen
  \bibfield  {author} {\bibinfo {author} {\bibfnamefont {F.~J.}\ \bibnamefont
  {Burnell}}, \bibinfo {author} {\bibfnamefont {S.~H.}\ \bibnamefont {Simon}},\
  and\ \bibinfo {author} {\bibfnamefont {J.~K.}\ \bibnamefont {Slingerland}},\
  }\href {https://doi.org/10.1103/PhysRevB.84.125434} {\bibfield  {journal}
  {\bibinfo  {journal} {Phys. Rev. B}\ }\textbf {\bibinfo {volume} {84}},\
  \bibinfo {pages} {125434} (\bibinfo {year} {2011})}\BibitemShut {NoStop}%
\bibitem [{\citenamefont {Burnell}(2018)}]{Burnell_2018}%
  \BibitemOpen
  \bibfield  {author} {\bibinfo {author} {\bibfnamefont {F.}~\bibnamefont
  {Burnell}},\ }\href
  {https://doi.org/10.1146/annurev-conmatphys-033117-054154} {\bibfield
  {journal} {\bibinfo  {journal} {Annu. Rev. Condens. Matter Phys.}\ }\textbf
  {\bibinfo {volume} {9}},\ \bibinfo {pages} {307} (\bibinfo {year}
  {2018})}\BibitemShut {NoStop}%
\bibitem [{\citenamefont {Trebst}\ \emph {et~al.}(2007)\citenamefont {Trebst},
  \citenamefont {Werner}, \citenamefont {Troyer}, \citenamefont {Shtengel},\
  and\ \citenamefont {Nayak}}]{Trebst_2007}%
  \BibitemOpen
  \bibfield  {author} {\bibinfo {author} {\bibfnamefont {S.}~\bibnamefont
  {Trebst}}, \bibinfo {author} {\bibfnamefont {P.}~\bibnamefont {Werner}},
  \bibinfo {author} {\bibfnamefont {M.}~\bibnamefont {Troyer}}, \bibinfo
  {author} {\bibfnamefont {K.}~\bibnamefont {Shtengel}},\ and\ \bibinfo
  {author} {\bibfnamefont {C.}~\bibnamefont {Nayak}},\ }\href
  {https://doi.org/10.1103/PhysRevLett.98.070602} {\bibfield  {journal}
  {\bibinfo  {journal} {Phys. Rev. Lett.}\ }\textbf {\bibinfo {volume} {98}},\
  \bibinfo {pages} {070602} (\bibinfo {year} {2007})}\BibitemShut {NoStop}%
\bibitem [{\citenamefont {Hamma}\ and\ \citenamefont
  {Lidar}(2008)}]{Hamma_2008_b}%
  \BibitemOpen
  \bibfield  {author} {\bibinfo {author} {\bibfnamefont {A.}~\bibnamefont
  {Hamma}}\ and\ \bibinfo {author} {\bibfnamefont {D.~A.}\ \bibnamefont
  {Lidar}},\ }\href {https://doi.org/10.1103/PhysRevLett.100.030502} {\bibfield
   {journal} {\bibinfo  {journal} {Phys. Rev. Lett.}\ }\textbf {\bibinfo
  {volume} {100}},\ \bibinfo {pages} {030502} (\bibinfo {year}
  {2008})}\BibitemShut {NoStop}%
\bibitem [{\citenamefont {Yu}\ \emph {et~al.}(2008)\citenamefont {Yu},
  \citenamefont {Kou},\ and\ \citenamefont {Wen}}]{Yu_2008}%
  \BibitemOpen
  \bibfield  {author} {\bibinfo {author} {\bibfnamefont {J.}~\bibnamefont
  {Yu}}, \bibinfo {author} {\bibfnamefont {S.-P.}\ \bibnamefont {Kou}},\ and\
  \bibinfo {author} {\bibfnamefont {X.-G.}\ \bibnamefont {Wen}},\ }\href
  {https://doi.org/10.1209/0295-5075/84/17004} {\bibfield  {journal} {\bibinfo
  {journal} {Eur. Phys. Lett.}\ }\textbf {\bibinfo {volume} {84}},\ \bibinfo
  {pages} {17004} (\bibinfo {year} {2008})}\BibitemShut {NoStop}%
\bibitem [{\citenamefont {Vidal}\ \emph
  {et~al.}(2009{\natexlab{a}})\citenamefont {Vidal}, \citenamefont {Dusuel},\
  and\ \citenamefont {Schmidt}}]{Vidal_2009}%
  \BibitemOpen
  \bibfield  {author} {\bibinfo {author} {\bibfnamefont {J.}~\bibnamefont
  {Vidal}}, \bibinfo {author} {\bibfnamefont {S.}~\bibnamefont {Dusuel}},\ and\
  \bibinfo {author} {\bibfnamefont {K.~P.}\ \bibnamefont {Schmidt}},\ }\href
  {https://doi.org/10.1103/PhysRevB.79.033109} {\bibfield  {journal} {\bibinfo
  {journal} {Phys. Rev. B}\ }\textbf {\bibinfo {volume} {79}},\ \bibinfo
  {pages} {033109} (\bibinfo {year} {2009}{\natexlab{a}})}\BibitemShut
  {NoStop}%
\bibitem [{\citenamefont {Vidal}\ \emph
  {et~al.}(2009{\natexlab{b}})\citenamefont {Vidal}, \citenamefont {Thomale},
  \citenamefont {Schmidt},\ and\ \citenamefont {Dusuel}}]{Vidal_2011}%
  \BibitemOpen
  \bibfield  {author} {\bibinfo {author} {\bibfnamefont {J.}~\bibnamefont
  {Vidal}}, \bibinfo {author} {\bibfnamefont {R.}~\bibnamefont {Thomale}},
  \bibinfo {author} {\bibfnamefont {K.~P.}\ \bibnamefont {Schmidt}},\ and\
  \bibinfo {author} {\bibfnamefont {S.}~\bibnamefont {Dusuel}},\ }\href
  {https://doi.org/10.1103/PhysRevB.80.081104} {\bibfield  {journal} {\bibinfo
  {journal} {Phys. Rev. B}\ }\textbf {\bibinfo {volume} {80}},\ \bibinfo
  {pages} {081104} (\bibinfo {year} {2009}{\natexlab{b}})}\BibitemShut
  {NoStop}%
\bibitem [{\citenamefont {Dusuel}\ \emph
  {et~al.}(2010{\natexlab{a}})\citenamefont {Dusuel}, \citenamefont {Kamfor},
  \citenamefont {Schmidt}, \citenamefont {Thomale},\ and\ \citenamefont
  {Vidal}}]{Dusuel_2009}%
  \BibitemOpen
  \bibfield  {author} {\bibinfo {author} {\bibfnamefont {S.}~\bibnamefont
  {Dusuel}}, \bibinfo {author} {\bibfnamefont {M.}~\bibnamefont {Kamfor}},
  \bibinfo {author} {\bibfnamefont {K.~P.}\ \bibnamefont {Schmidt}}, \bibinfo
  {author} {\bibfnamefont {R.}~\bibnamefont {Thomale}},\ and\ \bibinfo {author}
  {\bibfnamefont {J.}~\bibnamefont {Vidal}},\ }\href
  {https://doi.org/10.1103/PhysRevB.81.064412} {\bibfield  {journal} {\bibinfo
  {journal} {Phys. Rev. B}\ }\textbf {\bibinfo {volume} {81}},\ \bibinfo
  {pages} {064412} (\bibinfo {year} {2010}{\natexlab{a}})}\BibitemShut
  {NoStop}%
\bibitem [{\citenamefont {Tupitsyn}\ \emph {et~al.}(2010)\citenamefont
  {Tupitsyn}, \citenamefont {Kitaev}, \citenamefont {Prokof'ev},\ and\
  \citenamefont {Stamp}}]{Tupitsyn_2010}%
  \BibitemOpen
  \bibfield  {author} {\bibinfo {author} {\bibfnamefont {I.~S.}\ \bibnamefont
  {Tupitsyn}}, \bibinfo {author} {\bibfnamefont {A.}~\bibnamefont {Kitaev}},
  \bibinfo {author} {\bibfnamefont {N.~V.}\ \bibnamefont {Prokof'ev}},\ and\
  \bibinfo {author} {\bibfnamefont {P.~C.~E.}\ \bibnamefont {Stamp}},\ }\href
  {https://doi.org/10.1103/PhysRevB.82.085114} {\bibfield  {journal} {\bibinfo
  {journal} {Phys. Rev. B}\ }\textbf {\bibinfo {volume} {82}},\ \bibinfo
  {pages} {085114} (\bibinfo {year} {2010})}\BibitemShut {NoStop}%
\bibitem [{\citenamefont {Wu}\ \emph {et~al.}(2012)\citenamefont {Wu},
  \citenamefont {Deng},\ and\ \citenamefont {Prokof'ev}}]{Wu_2012}%
  \BibitemOpen
  \bibfield  {author} {\bibinfo {author} {\bibfnamefont {F.}~\bibnamefont
  {Wu}}, \bibinfo {author} {\bibfnamefont {Y.}~\bibnamefont {Deng}},\ and\
  \bibinfo {author} {\bibfnamefont {N.}~\bibnamefont {Prokof'ev}},\ }\href
  {https://doi.org/10.1103/PhysRevB.85.195104} {\bibfield  {journal} {\bibinfo
  {journal} {Phys. Rev. B}\ }\textbf {\bibinfo {volume} {85}},\ \bibinfo
  {pages} {195104} (\bibinfo {year} {2012})}\BibitemShut {NoStop}%
\bibitem [{\citenamefont {Dusuel}\ \emph {et~al.}(2011)\citenamefont {Dusuel},
  \citenamefont {Kamfor}, \citenamefont {Or\'{u}s}, \citenamefont {Schmidt},\
  and\ \citenamefont {Vidal}}]{Dusuel_2011}%
  \BibitemOpen
  \bibfield  {author} {\bibinfo {author} {\bibfnamefont {S.}~\bibnamefont
  {Dusuel}}, \bibinfo {author} {\bibfnamefont {M.}~\bibnamefont {Kamfor}},
  \bibinfo {author} {\bibfnamefont {R.}~\bibnamefont {Or\'{u}s}}, \bibinfo
  {author} {\bibfnamefont {K.~P.}\ \bibnamefont {Schmidt}},\ and\ \bibinfo
  {author} {\bibfnamefont {J.}~\bibnamefont {Vidal}},\ }\href
  {https://doi.org/10.1103/PhysRevLett.106.107203} {\bibfield  {journal}
  {\bibinfo  {journal} {Phys. Rev. Lett.}\ }\textbf {\bibinfo {volume} {106}},\
  \bibinfo {pages} {107203} (\bibinfo {year} {2011})}\BibitemShut {NoStop}%
\bibitem [{\citenamefont {Schmidt}(2013)}]{Schmidt_2013}%
  \BibitemOpen
  \bibfield  {author} {\bibinfo {author} {\bibfnamefont {K.~P.}\ \bibnamefont
  {Schmidt}},\ }\href {https://doi.org/10.1103/PhysRevB.88.035118} {\bibfield
  {journal} {\bibinfo  {journal} {Phys. Rev. B}\ }\textbf {\bibinfo {volume}
  {88}},\ \bibinfo {pages} {035118} (\bibinfo {year} {2013})}\BibitemShut
  {NoStop}%
\bibitem [{\citenamefont {Jahromi}\ \emph {et~al.}(2013)\citenamefont
  {Jahromi}, \citenamefont {Kargarian}, \citenamefont {Masoudi},\ and\
  \citenamefont {Schmidt}}]{Jahromi_2013}%
  \BibitemOpen
  \bibfield  {author} {\bibinfo {author} {\bibfnamefont {S.~S.}\ \bibnamefont
  {Jahromi}}, \bibinfo {author} {\bibfnamefont {M.}~\bibnamefont {Kargarian}},
  \bibinfo {author} {\bibfnamefont {S.~F.}\ \bibnamefont {Masoudi}},\ and\
  \bibinfo {author} {\bibfnamefont {K.~P.}\ \bibnamefont {Schmidt}},\ }\href
  {https://doi.org/10.1103/PhysRevB.87.094413} {\bibfield  {journal} {\bibinfo
  {journal} {Phys. Rev. B}\ }\textbf {\bibinfo {volume} {87}},\ \bibinfo
  {pages} {094413} (\bibinfo {year} {2013})}\BibitemShut {NoStop}%
\bibitem [{\citenamefont {Morampudi}\ \emph {et~al.}(2014)\citenamefont
  {Morampudi}, \citenamefont {von Keyserlingk},\ and\ \citenamefont
  {Pollmann}}]{Morampudi_2014}%
  \BibitemOpen
  \bibfield  {author} {\bibinfo {author} {\bibfnamefont {S.~C.}\ \bibnamefont
  {Morampudi}}, \bibinfo {author} {\bibfnamefont {C.}~\bibnamefont {von
  Keyserlingk}},\ and\ \bibinfo {author} {\bibfnamefont {F.}~\bibnamefont
  {Pollmann}},\ }\href {https://doi.org/10.1103/PhysRevB.90.035117} {\bibfield
  {journal} {\bibinfo  {journal} {Phys. Rev. B}\ }\textbf {\bibinfo {volume}
  {90}},\ \bibinfo {pages} {035117} (\bibinfo {year} {2014})}\BibitemShut
  {NoStop}%
\bibitem [{\citenamefont {Schulz}\ and\ \citenamefont
  {Burnell}(2016)}]{Schulz_2016}%
  \BibitemOpen
  \bibfield  {author} {\bibinfo {author} {\bibfnamefont {M.~D.}\ \bibnamefont
  {Schulz}}\ and\ \bibinfo {author} {\bibfnamefont {F.~J.}\ \bibnamefont
  {Burnell}},\ }\href {https://doi.org/10.1103/PhysRevB.94.165110} {\bibfield
  {journal} {\bibinfo  {journal} {Phys. Rev. B}\ }\textbf {\bibinfo {volume}
  {94}},\ \bibinfo {pages} {165110} (\bibinfo {year} {2016})}\BibitemShut
  {NoStop}%
\bibitem [{\citenamefont {Zhang}\ \emph {et~al.}(2017)\citenamefont {Zhang},
  \citenamefont {Melko},\ and\ \citenamefont {Kim}}]{Zhang_2017}%
  \BibitemOpen
  \bibfield  {author} {\bibinfo {author} {\bibfnamefont {Y.}~\bibnamefont
  {Zhang}}, \bibinfo {author} {\bibfnamefont {R.~G.}\ \bibnamefont {Melko}},\
  and\ \bibinfo {author} {\bibfnamefont {E.-A.}\ \bibnamefont {Kim}},\ }\href
  {https://doi.org/10.1103/PhysRevB.96.245119} {\bibfield  {journal} {\bibinfo
  {journal} {Phys. Rev. B}\ }\textbf {\bibinfo {volume} {96}},\ \bibinfo
  {pages} {245119} (\bibinfo {year} {2017})}\BibitemShut {NoStop}%
\bibitem [{\citenamefont {Vanderstraeten}\ \emph {et~al.}(2017)\citenamefont
  {Vanderstraeten}, \citenamefont {Mari\"en}, \citenamefont {Haegeman},
  \citenamefont {Schuch}, \citenamefont {Vidal},\ and\ \citenamefont
  {Verstraete}}]{Vanderstraeten_2017}%
  \BibitemOpen
  \bibfield  {author} {\bibinfo {author} {\bibfnamefont {L.}~\bibnamefont
  {Vanderstraeten}}, \bibinfo {author} {\bibfnamefont {M.}~\bibnamefont
  {Mari\"en}}, \bibinfo {author} {\bibfnamefont {J.}~\bibnamefont {Haegeman}},
  \bibinfo {author} {\bibfnamefont {N.}~\bibnamefont {Schuch}}, \bibinfo
  {author} {\bibfnamefont {J.}~\bibnamefont {Vidal}},\ and\ \bibinfo {author}
  {\bibfnamefont {F.}~\bibnamefont {Verstraete}},\ }\href
  {https://doi.org/10.1103/PhysRevLett.119.070401} {\bibfield  {journal}
  {\bibinfo  {journal} {Phys. Rev. Lett.}\ }\textbf {\bibinfo {volume} {119}},\
  \bibinfo {pages} {070401} (\bibinfo {year} {2017})}\BibitemShut {NoStop}%
\bibitem [{\citenamefont {Wen}(2000)}]{Wen_2010}%
  \BibitemOpen
  \bibfield  {author} {\bibinfo {author} {\bibfnamefont {X.-G.}\ \bibnamefont
  {Wen}},\ }\href {https://doi.org/10.1103/PhysRevLett.84.3950} {\bibfield
  {journal} {\bibinfo  {journal} {Phys. Rev. Lett.}\ }\textbf {\bibinfo
  {volume} {84}},\ \bibinfo {pages} {3950} (\bibinfo {year}
  {2000})}\BibitemShut {NoStop}%
\bibitem [{\citenamefont {Barkeshli}\ and\ \citenamefont
  {Wen}(2010)}]{Barkeshli_2010}%
  \BibitemOpen
  \bibfield  {author} {\bibinfo {author} {\bibfnamefont {M.}~\bibnamefont
  {Barkeshli}}\ and\ \bibinfo {author} {\bibfnamefont {X.-G.}\ \bibnamefont
  {Wen}},\ }\href {https://doi.org/10.1103/PhysRevLett.105.216804} {\bibfield
  {journal} {\bibinfo  {journal} {Phys. Rev. Lett.}\ }\textbf {\bibinfo
  {volume} {105}},\ \bibinfo {pages} {216804} (\bibinfo {year}
  {2010})}\BibitemShut {NoStop}%
\bibitem [{\citenamefont {M\"oller}\ \emph {et~al.}(2014)\citenamefont
  {M\"oller}, \citenamefont {Hormozi}, \citenamefont {Slingerland},\ and\
  \citenamefont {Simon}}]{Moeller_2014}%
  \BibitemOpen
  \bibfield  {author} {\bibinfo {author} {\bibfnamefont {G.}~\bibnamefont
  {M\"oller}}, \bibinfo {author} {\bibfnamefont {L.}~\bibnamefont {Hormozi}},
  \bibinfo {author} {\bibfnamefont {J.}~\bibnamefont {Slingerland}},\ and\
  \bibinfo {author} {\bibfnamefont {S.~H.}\ \bibnamefont {Simon}},\ }\href
  {https://doi.org/10.1103/PhysRevB.90.235101} {\bibfield  {journal} {\bibinfo
  {journal} {Phys. Rev. B}\ }\textbf {\bibinfo {volume} {90}},\ \bibinfo
  {pages} {235101} (\bibinfo {year} {2014})}\BibitemShut {NoStop}%
\bibitem [{\citenamefont {Bombin}\ and\ \citenamefont
  {Martin-Delgado}(2008)}]{Bombin_2008}%
  \BibitemOpen
  \bibfield  {author} {\bibinfo {author} {\bibfnamefont {H.}~\bibnamefont
  {Bombin}}\ and\ \bibinfo {author} {\bibfnamefont {M.~A.}\ \bibnamefont
  {Martin-Delgado}},\ }\href {https://doi.org/10.1103/PhysRevB.78.115421}
  {\bibfield  {journal} {\bibinfo  {journal} {Phys. Rev. B}\ }\textbf {\bibinfo
  {volume} {78}},\ \bibinfo {pages} {115421} (\bibinfo {year}
  {2008})}\BibitemShut {NoStop}%
\bibitem [{\citenamefont {Fuji}(2019)}]{Fujii_2019}%
  \BibitemOpen
  \bibfield  {author} {\bibinfo {author} {\bibfnamefont {Y.}~\bibnamefont
  {Fuji}},\ }\href {https://doi.org/10.1103/PhysRevB.100.235115} {\bibfield
  {journal} {\bibinfo  {journal} {Phys. Rev. B}\ }\textbf {\bibinfo {volume}
  {100}},\ \bibinfo {pages} {235115} (\bibinfo {year} {2019})}\BibitemShut
  {NoStop}%
\bibitem [{\citenamefont {Wiedmann}\ \emph {et~al.}(2020)\citenamefont
  {Wiedmann}, \citenamefont {Lenke}, \citenamefont {Walther}, \citenamefont
  {M\"uhlhauser},\ and\ \citenamefont {Schmidt}}]{Wiedmann_2020}%
  \BibitemOpen
  \bibfield  {author} {\bibinfo {author} {\bibfnamefont {R.}~\bibnamefont
  {Wiedmann}}, \bibinfo {author} {\bibfnamefont {L.}~\bibnamefont {Lenke}},
  \bibinfo {author} {\bibfnamefont {M.~R.}\ \bibnamefont {Walther}}, \bibinfo
  {author} {\bibfnamefont {M.}~\bibnamefont {M\"uhlhauser}},\ and\ \bibinfo
  {author} {\bibfnamefont {K.~P.}\ \bibnamefont {Schmidt}},\ }\href
  {https://doi.org/10.1103/PhysRevB.102.214422} {\bibfield  {journal} {\bibinfo
   {journal} {Phys. Rev. B}\ }\textbf {\bibinfo {volume} {102}},\ \bibinfo
  {pages} {214422} (\bibinfo {year} {2020})}\BibitemShut {NoStop}%
\bibitem [{\citenamefont {Vidal}\ \emph {et~al.}(2008)\citenamefont {Vidal},
  \citenamefont {Schmidt},\ and\ \citenamefont {Dusuel}}]{Vidal_2008}%
  \BibitemOpen
  \bibfield  {author} {\bibinfo {author} {\bibfnamefont {J.}~\bibnamefont
  {Vidal}}, \bibinfo {author} {\bibfnamefont {K.~P.}\ \bibnamefont {Schmidt}},\
  and\ \bibinfo {author} {\bibfnamefont {S.}~\bibnamefont {Dusuel}},\ }\href
  {https://doi.org/10.1103/PhysRevB.78.245121} {\bibfield  {journal} {\bibinfo
  {journal} {Phys. Rev. B}\ }\textbf {\bibinfo {volume} {78}},\ \bibinfo
  {pages} {245121} (\bibinfo {year} {2008})}\BibitemShut {NoStop}%
\bibitem [{\citenamefont {Takahashi}(1977)}]{Takahashi_1977}%
  \BibitemOpen
  \bibfield  {author} {\bibinfo {author} {\bibfnamefont {M.}~\bibnamefont
  {Takahashi}},\ }\href {https://doi.org/10.1088/0022-3719/10/8/031} {\bibfield
   {journal} {\bibinfo  {journal} {Journal of Physics C: Solid State Physics}\
  }\textbf {\bibinfo {volume} {10}},\ \bibinfo {pages} {1289} (\bibinfo {year}
  {1977})}\BibitemShut {NoStop}%
\bibitem [{\citenamefont {Klagges}\ and\ \citenamefont
  {Schmidt}(2012)}]{Klagges_2012}%
  \BibitemOpen
  \bibfield  {author} {\bibinfo {author} {\bibfnamefont {D.}~\bibnamefont
  {Klagges}}\ and\ \bibinfo {author} {\bibfnamefont {K.~P.}\ \bibnamefont
  {Schmidt}},\ }\href {https://doi.org/10.1103/PhysRevLett.108.230508}
  {\bibfield  {journal} {\bibinfo  {journal} {Phys. Rev. Lett.}\ }\textbf
  {\bibinfo {volume} {108}},\ \bibinfo {pages} {230508} (\bibinfo {year}
  {2012})}\BibitemShut {NoStop}%
\bibitem [{\citenamefont {Castelnovo}\ and\ \citenamefont
  {Chamon}(2008)}]{Castelnovo_2008}%
  \BibitemOpen
  \bibfield  {author} {\bibinfo {author} {\bibfnamefont {C.}~\bibnamefont
  {Castelnovo}}\ and\ \bibinfo {author} {\bibfnamefont {C.}~\bibnamefont
  {Chamon}},\ }\href {https://doi.org/10.1103/PhysRevB.78.155120} {\bibfield
  {journal} {\bibinfo  {journal} {Phys. Rev. B}\ }\textbf {\bibinfo {volume}
  {78}},\ \bibinfo {pages} {155120} (\bibinfo {year} {2008})}\BibitemShut
  {NoStop}%
\bibitem [{\citenamefont {Schmidt}\ \emph {et~al.}(2008)\citenamefont
  {Schmidt}, \citenamefont {Dusuel},\ and\ \citenamefont
  {Vidal}}]{Schmidt_2008}%
  \BibitemOpen
  \bibfield  {author} {\bibinfo {author} {\bibfnamefont {K.~P.}\ \bibnamefont
  {Schmidt}}, \bibinfo {author} {\bibfnamefont {S.}~\bibnamefont {Dusuel}},\
  and\ \bibinfo {author} {\bibfnamefont {J.}~\bibnamefont {Vidal}},\ }\href
  {https://doi.org/10.1103/PhysRevLett.100.057208} {\bibfield  {journal}
  {\bibinfo  {journal} {Phys. Rev. Lett.}\ }\textbf {\bibinfo {volume} {100}},\
  \bibinfo {pages} {057208} (\bibinfo {year} {2008})}\BibitemShut {NoStop}%
\bibitem [{\citenamefont {Gover}(1994)}]{Gover_1994}%
  \BibitemOpen
  \bibfield  {author} {\bibinfo {author} {\bibfnamefont {M.}~\bibnamefont
  {Gover}},\ }\href
  {https://doi.org/https://doi.org/10.1016/0024-3795(94)90481-2} {\bibfield
  {journal} {\bibinfo  {journal} {Linear Algebra and its Applications}\
  }\textbf {\bibinfo {volume} {197-198}},\ \bibinfo {pages} {63} (\bibinfo
  {year} {1994})}\BibitemShut {NoStop}%
\bibitem [{\citenamefont {L\"{o}wdin}(1962)}]{Loewdin1962}%
  \BibitemOpen
  \bibfield  {author} {\bibinfo {author} {\bibfnamefont {P.-O.}\ \bibnamefont
  {L\"{o}wdin}},\ }\href {https://doi.org/10.1063/1.1724312} {\bibfield
  {journal} {\bibinfo  {journal} {Journal of Mathematical Physics}\ }\textbf
  {\bibinfo {volume} {3}},\ \bibinfo {pages} {969} (\bibinfo {year}
  {1962})}\BibitemShut {NoStop}%
\bibitem [{\citenamefont {Yao}\ and\ \citenamefont {Shi}(2000)}]{Yao2000}%
  \BibitemOpen
  \bibfield  {author} {\bibinfo {author} {\bibfnamefont {D.}~\bibnamefont
  {Yao}}\ and\ \bibinfo {author} {\bibfnamefont {J.}~\bibnamefont {Shi}},\
  }\href {https://doi.org/10.1119/1.19419} {\bibfield  {journal} {\bibinfo
  {journal} {American Journal of Physics}\ }\textbf {\bibinfo {volume} {68}},\
  \bibinfo {pages} {278} (\bibinfo {year} {2000})}\BibitemShut {NoStop}%
\bibitem [{\citenamefont {Kalis}\ \emph {et~al.}(2012)\citenamefont {Kalis},
  \citenamefont {Klagges}, \citenamefont {Or\'us},\ and\ \citenamefont
  {Schmidt}}]{Kalis_2012}%
  \BibitemOpen
  \bibfield  {author} {\bibinfo {author} {\bibfnamefont {H.}~\bibnamefont
  {Kalis}}, \bibinfo {author} {\bibfnamefont {D.}~\bibnamefont {Klagges}},
  \bibinfo {author} {\bibfnamefont {R.}~\bibnamefont {Or\'us}},\ and\ \bibinfo
  {author} {\bibfnamefont {K.~P.}\ \bibnamefont {Schmidt}},\ }\href
  {https://doi.org/10.1103/PhysRevA.86.022317} {\bibfield  {journal} {\bibinfo
  {journal} {Phys. Rev. A}\ }\textbf {\bibinfo {volume} {86}},\ \bibinfo
  {pages} {022317} (\bibinfo {year} {2012})}\BibitemShut {NoStop}%
\bibitem [{\citenamefont {Knetter}\ and\ \citenamefont
  {Uhrig}(2000)}]{Knetter_2000}%
  \BibitemOpen
  \bibfield  {author} {\bibinfo {author} {\bibfnamefont {C.}~\bibnamefont
  {Knetter}}\ and\ \bibinfo {author} {\bibfnamefont {G.~S.}\ \bibnamefont
  {Uhrig}},\ }\href {https://doi.org/10.1007/s100510050026} {\bibfield
  {journal} {\bibinfo  {journal} {Eur. Phys. J. B}\ }\textbf {\bibinfo {volume}
  {13}},\ \bibinfo {pages} {209} (\bibinfo {year} {2000})}\BibitemShut
  {NoStop}%
\bibitem [{\citenamefont {Knetter}\ \emph {et~al.}(2003)\citenamefont
  {Knetter}, \citenamefont {Schmidt},\ and\ \citenamefont
  {Uhrig}}]{Knetter2003}%
  \BibitemOpen
  \bibfield  {author} {\bibinfo {author} {\bibfnamefont {C.}~\bibnamefont
  {Knetter}}, \bibinfo {author} {\bibfnamefont {K.~P.}\ \bibnamefont
  {Schmidt}},\ and\ \bibinfo {author} {\bibfnamefont {G.~S.}\ \bibnamefont
  {Uhrig}},\ }\href {https://doi.org/10.1088/0305-4470/36/29/302} {\bibfield
  {journal} {\bibinfo  {journal} {Journal of Physics A: Mathematical and
  General}\ }\textbf {\bibinfo {volume} {36}},\ \bibinfo {pages} {7889}
  (\bibinfo {year} {2003})}\BibitemShut {NoStop}%
\bibitem [{\citenamefont {Guttmann}(1989)}]{Guttmann_1989}%
  \BibitemOpen
  \bibfield  {author} {\bibinfo {author} {\bibfnamefont {A.}~\bibnamefont
  {Guttmann}},\ }\bibinfo {title} {Phase transitions and critical phenomena}\
  (\bibinfo  {publisher} {Academic, New York},\ \bibinfo {year}
  {1989})\BibitemShut {NoStop}%
\bibitem [{\citenamefont {Pfeuty}\ and\ \citenamefont
  {Elliott}(1971)}]{Pfeuty_1971}%
  \BibitemOpen
  \bibfield  {author} {\bibinfo {author} {\bibfnamefont {P.}~\bibnamefont
  {Pfeuty}}\ and\ \bibinfo {author} {\bibfnamefont {R.~J.}\ \bibnamefont
  {Elliott}},\ }\href {https://doi.org/10.1088/0022-3719/4/15/024} {\bibfield
  {journal} {\bibinfo  {journal} {Journal of Physics C: Solid State Physics}\
  }\textbf {\bibinfo {volume} {4}},\ \bibinfo {pages} {2370} (\bibinfo {year}
  {1971})}\BibitemShut {NoStop}%
\bibitem [{\citenamefont {Kos}\ \emph {et~al.}(2016)\citenamefont {Kos},
  \citenamefont {Poland}, \citenamefont {Simmons-Duffin},\ and\ \citenamefont
  {Vichi}}]{Kos_2016}%
  \BibitemOpen
  \bibfield  {author} {\bibinfo {author} {\bibfnamefont {F.}~\bibnamefont
  {Kos}}, \bibinfo {author} {\bibfnamefont {D.}~\bibnamefont {Poland}},
  \bibinfo {author} {\bibfnamefont {D.}~\bibnamefont {Simmons-Duffin}},\ and\
  \bibinfo {author} {\bibfnamefont {A.}~\bibnamefont {Vichi}},\ }\href
  {https://doi.org/10.1007/JHEP08(2016)036} {\bibfield  {journal} {\bibinfo
  {journal} {Journal of High Energy Physics}\ }\textbf {\bibinfo {volume}
  {2016}},\ \bibinfo {pages} {36} (\bibinfo {year} {2016})}\BibitemShut
  {NoStop}%
\bibitem [{\citenamefont {He}\ \emph {et~al.}(1990)\citenamefont {He},
  \citenamefont {Hamer},\ and\ \citenamefont {Oitmaa}}]{He_1990}%
  \BibitemOpen
  \bibfield  {author} {\bibinfo {author} {\bibfnamefont {H.~X.}\ \bibnamefont
  {He}}, \bibinfo {author} {\bibfnamefont {C.~J.}\ \bibnamefont {Hamer}},\ and\
  \bibinfo {author} {\bibfnamefont {J.}~\bibnamefont {Oitmaa}},\ }\href
  {https://doi.org/10.1088/0305-4470/23/10/018} {\bibfield  {journal} {\bibinfo
   {journal} {Journal of Physics A: Mathematical and General}\ }\textbf
  {\bibinfo {volume} {23}},\ \bibinfo {pages} {1775} (\bibinfo {year}
  {1990})}\BibitemShut {NoStop}%
\bibitem [{\citenamefont {Griffin}\ and\ \citenamefont
  {Bartlett}(2008)}]{Griffin_2008}%
  \BibitemOpen
  \bibfield  {author} {\bibinfo {author} {\bibfnamefont {T.}~\bibnamefont
  {Griffin}}\ and\ \bibinfo {author} {\bibfnamefont {S.~D.}\ \bibnamefont
  {Bartlett}},\ }\href {https://doi.org/10.1103/PhysRevA.78.062306} {\bibfield
  {journal} {\bibinfo  {journal} {Phys. Rev. A}\ }\textbf {\bibinfo {volume}
  {78}},\ \bibinfo {pages} {062306} (\bibinfo {year} {2008})}\BibitemShut
  {NoStop}%
\bibitem [{\citenamefont {Dusuel}\ \emph
  {et~al.}(2010{\natexlab{b}})\citenamefont {Dusuel}, \citenamefont {Kamfor},
  \citenamefont {Schmidt}, \citenamefont {Thomale},\ and\ \citenamefont
  {Vidal}}]{Dusuel2010}%
  \BibitemOpen
  \bibfield  {author} {\bibinfo {author} {\bibfnamefont {S.}~\bibnamefont
  {Dusuel}}, \bibinfo {author} {\bibfnamefont {M.}~\bibnamefont {Kamfor}},
  \bibinfo {author} {\bibfnamefont {K.~P.}\ \bibnamefont {Schmidt}}, \bibinfo
  {author} {\bibfnamefont {R.}~\bibnamefont {Thomale}},\ and\ \bibinfo {author}
  {\bibfnamefont {J.}~\bibnamefont {Vidal}},\ }\href
  {https://doi.org/10.1103/PhysRevB.81.064412} {\bibfield  {journal} {\bibinfo
  {journal} {Phys. Rev. B}\ }\textbf {\bibinfo {volume} {81}},\ \bibinfo
  {pages} {064412} (\bibinfo {year} {2010}{\natexlab{b}})}\BibitemShut
  {NoStop}%
\bibitem [{\citenamefont {Coester}\ and\ \citenamefont
  {Schmidt}(2015)}]{coester2015optimizing}%
  \BibitemOpen
  \bibfield  {author} {\bibinfo {author} {\bibfnamefont {K.}~\bibnamefont
  {Coester}}\ and\ \bibinfo {author} {\bibfnamefont {K.~P.}\ \bibnamefont
  {Schmidt}},\ }\href {https://doi.org/10.1103/physreve.92.022118} {\bibfield
  {journal} {\bibinfo  {journal} {Physical Review E}\ }\textbf {\bibinfo
  {volume} {92}},\ \bibinfo {pages} {022118} (\bibinfo {year}
  {2015})}\BibitemShut {NoStop}%
\end{thebibliography}
\end{document}